\documentclass[vecphys]{svmult}

\usepackage{makeidx}         
\usepackage{graphicx}        
\usepackage{multicol}        
\usepackage[bottom]{footmisc}

\makeindex             

\begin{document}

\title*{Statistical Features of Earthquake Temporal Occurrence}
\author{
\'Alvaro Corral
}
\institute{
Departament de F\'{\i}sica,
Facultat de Ci\`encies,
Universitat Aut\`onoma de Barcelona,
E-08193 Bellaterra,
Spain,
\texttt{Alvaro.Corral@uab.es}
}
%
%
\maketitle

\setcounter{page}{195}

{\it The physics of an earthquake} is a subject 
with many unknowns.
It is true that we have a good understanding
of the propagation of seismic waves through the Earth 
and that given a large set of seismographic 
records we are able to reconstruct {\it a posteriori}
the history of the fault rupture (the origin of the waves).
However, when we consider the physical processes which 
lead to the initiation of a rupture with a subsequent slip 
and its growth
through a fault system to give rise to an earthquake, 
then our knowledge is really limited.
Not only the friction law and the rupture evolution rules are
largely unknown, but the role of many other processes such as 
plasticity, fluid migration, chemical reactions, etc., and
the couplings between them, remain unclear
\cite{Mulargia_Geller,Rundle_review}.

On the other hand, one may wonder about
{\it the physics of many earthquakes}.
How do the collective properties of the set
defined by all earthquakes in a given region,
or better, in the whole world, 
emerge from the physics of individual earthquakes?
How does seismicity, which is the structure formed
by all earthquakes, depend on its elementary constituents
--the earthquakes?
And which are these properties? 
Which kind of dynamical process does seismicity
constitute?
It may be that these collective properties 
are largely independent on the physics of the individual earthquakes,
in the same way that many of the properties of a gas or a solid 
do not depend on the constitution of 
its elementary units --the atoms
(for a broad range of temperatures it doesn't matter if we have 
atoms, with its complicated quantum structure,
or microscopic marbles).
It is natural then to consider that 
the physics of many earthquakes has to be studied 
with a different approach than the physics of one
earthquake, and in this sense we can consider
the use of statistical physics not only appropriate 
but necessary to understand 
the collective properties of earthquakes.

Here, we provide a summary of recent work on the statistics 
of the temporal properties of seismicity, 
considering the phenomenon as a whole and
with the goal of looking for general laws.
We show the fulfillment of a scaling law 
for recurrence-time distributions, which becomes
universal for stationary seismicity and for
aftershock sequences which are transformed into
stationary processes by means of a nonlinear rescaling
of time.
The existence of a decreasing power-law regime 
in the distributions has paradoxical 
consequences on the time evolution of the earthquake hazard
and on the expected time of occurrence of an incoming event,
as we will see.
On the other hand, the scaling law for recurrence times is equivalent to 
the invariance of seismicity under renormalization-group-like transformations,
for which the role of correlations between recurrence times and 
magnitudes is essential.
Finally, we relate the recurrence-time densities studied here
with the method previously introduced by Bak et al.
\cite{Bak.2002}.

\section{The Gutenberg-Richter Law and the Omori Law}

Traditionally, the knowledge of seismicity has been 
limited to a few phenomenological laws, the most important
being the Gutenberg-Richter (GR) law                             \index{Gutenberg-Richter law}
and the Omori law.
The GR law determines that,
for a certain region,
the number of earthquakes in a long period of time 
decreases exponentially with the magnitude; 
to be concrete, $N(M_c) \propto 10^{-b M_c}$,
where $N(M_c)$ is the number of earthquakes with magnitude
$M$ greater or equal than a threshold value $M_c$,
and the $b-$value is a constant usually close to one
\cite{Kagan_physD,Turcotte_book,Utsu_GR,Utsu_handbook}.

If we introduce the seismic rate,                                 \index{seismic rate}
$r(t,M_c)$,
defined as the number of earthquakes with $M \ge M_c$ per unit time 
in a time interval around $t$,
then, the GR relation can be expressed in terms of the mean
seismic rate,                                                      \index{mean seismic rate}
$R(M_c)$, as
\begin{equation}
R(M_c) \equiv \langle r(t,M_c) \rangle =
\frac 1 T \int_0^T r(t,M_c) dt = \frac{N(M_c)} {T}
=R_0 10^{-b M_c},
\end{equation}
where $T$ is the total time under consideration and 
$R_0$ is the (hypothetical) mean rate in the region for $M_c=0$
(its dependence, as well as that of other parameters,
on the region selected for study is implicit
and will not be indicated when it is superfluous).
In fact, the GR law must be
understood as a probabilistic law, and then we conclude that
earthquake magnitude follows an exponential distribution, this is,
$\mbox{Prob}[M\ge M_c] = N(M_c) / \mathcal{N} \propto e^{-\ln 10 \, b M_c}$, 
with $\mathcal{N}$ the total number of earthquakes, of any magnitude.
Due to the properties of the exponential distribution,
the derivative of $\mbox{Prob}[M\ge M_c] $, 
which is the probability density (with a minus sign),
is also an exponential.

In terms of the seismic moment or of the dissipated
energy, which are increasing exponential functions
of the magnitude, the GR law transforms into a power-law distribution,       \index{power-law distribution}
the usual signature of scale invariance.
This means that earthquakes have no characteristic size of
occurrence, if we take the seismic moment or the energy
as more appropriate measures of earthquake size 
than the magnitude \cite{Kagan_physD}.

The Omori law (in its modified form)                      \index{Omori law} \index{modified Omori law}
states that after a strong earthquake, which
is called {\it mainshock}, the seismic rate
for events with $M\ge M_c$
in a certain region around the mainshock 
increases abruptly and then 
decays in time essentially as a power law; 
more precisely,
\begin{equation}
   r(t,M_c) = \frac {r_0(M_c)} {(1+t/c)^p}
\label{omori}
\end{equation}
where $t$ is the time measured from the mainshock,
$r_0(M_c)$ is the maximum rate for $M\ge M_c$, 
which coincides with the rate 
immediately after the mainshock, i.e.,  
$r(t=0,M_c)=r_0(M_c)$,
$c$ is a short-time constant (of the order of hours
or a few days), which describes the deviation from
a pure power law right after the mainshock,
and the exponent $p$ is usually close to 1.
In fact, $c$ depends to a certain degree on $M_c$ and, together
with $r_0(M_c)$ and $p$, depends also on the mainshock magnitude
\cite{Reasenberg,Utsu_omori,Utsu_handbook}.

Nowadays it has been confirmed that the Omori law does not only apply to
strong earthquakes, but to any earthquake, with a productivity 
factor ($r_0$) for small earthquakes which is 
orders of magnitude smaller than for large events. 
In this way, the classification of earthquakes in mainshocks
and aftershocks turns out to be only relative, 
as we will have a cascade process
in which aftershocks become also mainshocks of secondary sequences
and so on.
When an aftershock happens to have a magnitude larger than the mainshock
a change of roles occur: 
the mainshock is considered a foreshock
and
the aftershock becomes the mainshock.
Also, the triggering of strong aftershocks may cause
that the overall seismic rate departs significantly from the
Omori law, as it happens in earthquake swarms.

In any case, the Omori law illustrates clearly the temporal
clustering of earthquakes,                                        \index{clustering}
for which events (aftershocks) tend to gather close
(in time) to a strong event (the mainshock), 
becoming more dilute as time from the mainshocks grows.
In addition, 
the fact that the seismic rate
decays essentially as a power law
means that the relaxation process
has no characteristic time, in opposition
to the usual situation in physics
(think for instance in radioactive decay).
Finally, 
the Omori law has a probabilistic interpretation,
as an all-return-time distribution,
measuring the probability that earthquakes occur
at a time $t$ after a mainshock.

\section{Recurrence-Time Distributions
and Scaling Laws}

One can go beyond the GR law and the Omori law
and wonder about 
the temporal properties of individual earthquakes
(from a statistical point of view), in particular about
the time interval between consecutive earthquakes.
In this case, it is necessary to assume that earthquakes are point
events in time, or at least that their temporal properties are well 
described by their initiation time. 
In contrast to the previous approaches,
this perspective has been much less studied and no general law
has been proposed; rather, the situation is confusing in
the literature, where claims range from nearly-periodic behavior for large earthquakes
to totally random occurrence of mainshocks
(see the citations at Refs. 
\cite{Corral_npg.2005,Corral_tectono}).
Furthermore, it can be argued that the times between consecutive earthquakes
depend strongly on the selection of the coordinates of the region under study
and the range of magnitudes selected (which change the sequence of events)
and therefore one is dealing with an ill-defined variable.
We will see that the existence of universal properties for these times 
invalidates this objection.

Following the point of view of Bak et al.,                                  \index{Bak}
we have addressed this problem
by considering seismicity as a phenomenon on its own.
In this way, we will not separate events into different kinds
(foreshocks, mainshocks, aftershocks, or microearthquakes, etc.),
nor divide the crust into provinces with different tectonic properties,
but will place all events and regions on the same footing;
in other words, we wonder about the very nature of seismicity
as a whole,
from a complex-system perspective,                                          \index{complex-system perspective}
in opposition to a reductionist approach
\cite{Bak.2002,Bak_book}.

This exposition will concentrate on the temporal properties of
seismicity, and their dependence with space and magnitude, 
but equally important are the spatial properties.
It turns out that all the aspects of seismicity
are closely related to each other and one cannot study them 
separately.
Although all the events are important, as they are the 
elementary constituents of seismicity, 
we will need to consider windows of observation
in space, time, and magnitude; of course, this is due to the 
incompleteness of seismic records but also to the fact that
the variation of the quantities we measure with the 
range of magnitudes selected or with the size of the spatial
region under study
will allow us to establish self-similar properties for seismicity.

Let us select an arbitrary region of the Earth, a temporal period,
and a minimum magnitude $M_c$, in such a way that only events
in this space-time-magnitude window are taken into account.
We can consider the resulting events as a point process in time,
disregarding the magnitude and the spatial degrees of freedom
(this is not arbitrary, as $M_c$ and the size of the region 
will be systematically varied later on), in this way we can order
the events in time, from $i=1$ to $N(M_c)$ and 
characterize each one only by its occurrence time, $t_i$.
From here we can define the recurrence time $\tau$                       \index{recurrence time}
(also called waiting time,                                               \index{waiting time}
interevent time, interoccurrence time, etc.)
as the time interval between consecutive events, i.e.,
$\tau_i\equiv t_i-t_{i-1}$. 
The mean recurrence time, $\langle \tau(M_c)\rangle$,
is obviously given by the inverse of the rate, $R^{-1}(M_c)$;
however, 
as the recurrence time is broadly distributed, 
the mean alone is a poor characterization of the process
and it is inevitable to work with the probability distribution
of recurrence times.
So, we compute the recurrence-time probability density                   \index{recurrence-time probability density}
as
\begin{equation}
D(\tau; M_c) = \frac{\mbox{Prob}[\tau < \mbox{ recurrence time } 
\le \tau +d\tau]}{d\tau},
\end{equation}
where $d\tau$ has to be small enough to allow $D$ to represent a continuous
function but large enough to contain enough data to be statistically
significant (note that the spatial dependence of $D$ is not indicated explicitly).

\subsection{Scaling Laws for Recurrence-Time Distributions}

\begin{figure}
\centering
\includegraphics[height=4.1cm]{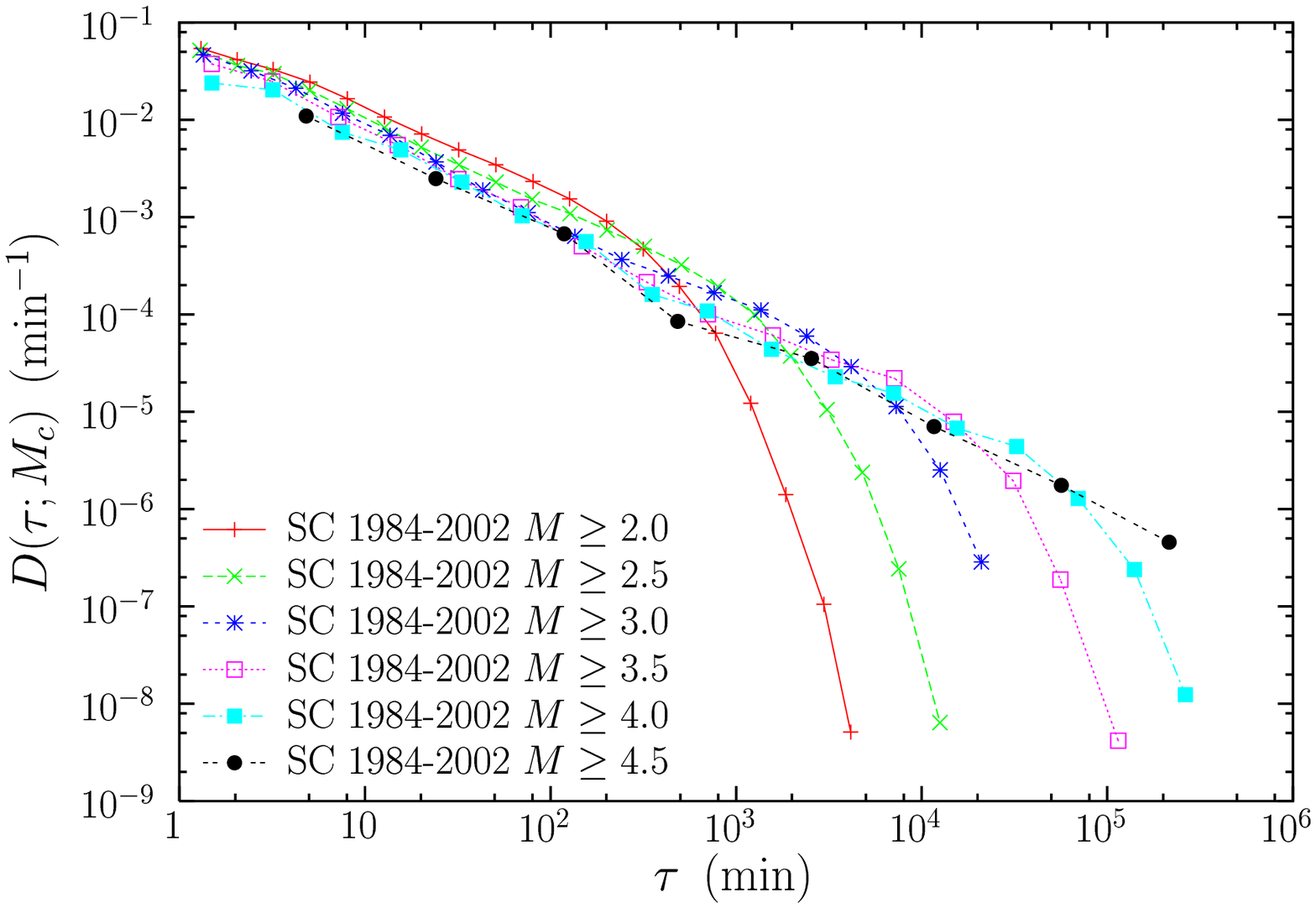}
\includegraphics[height=4.1cm]{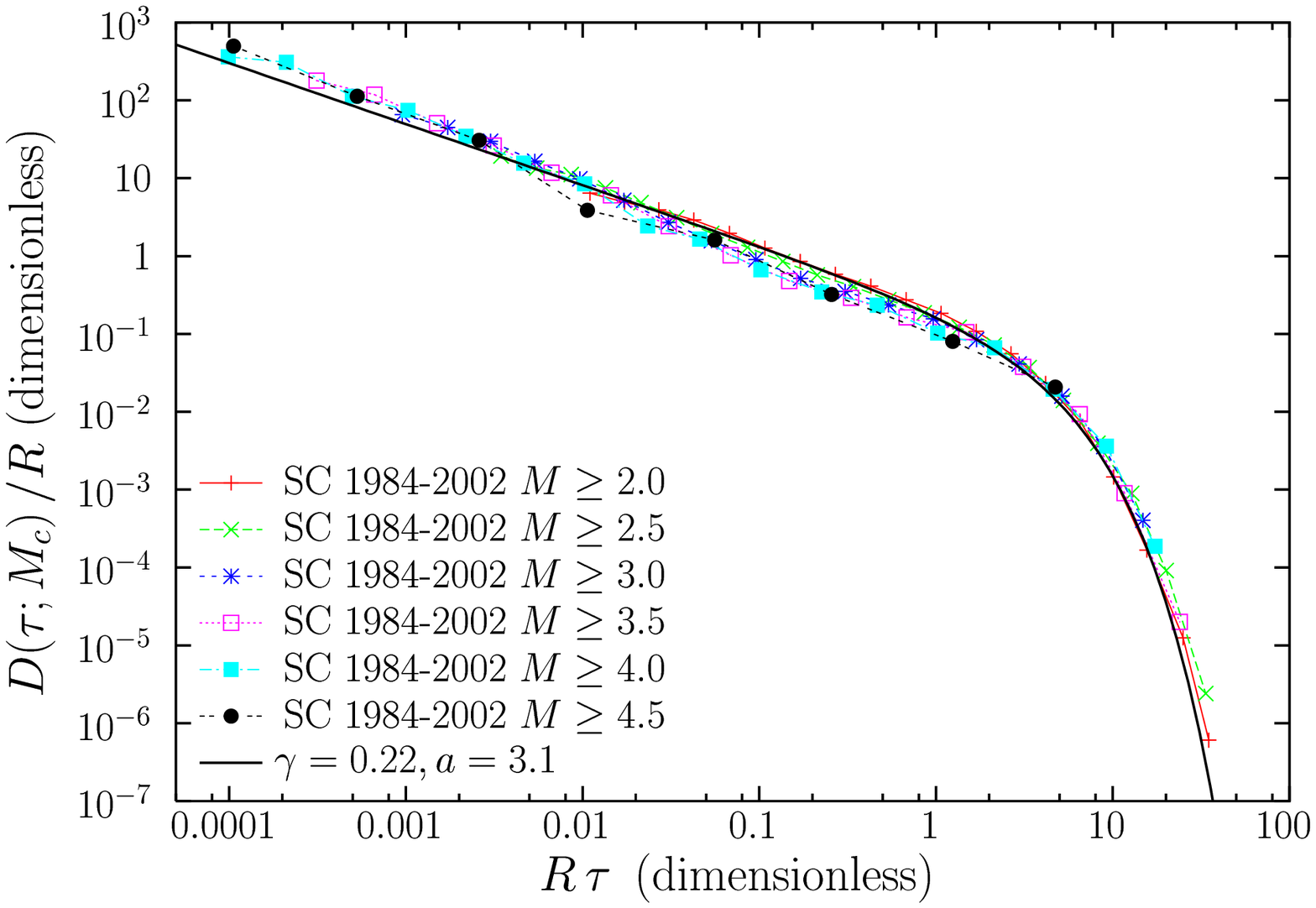}
\caption{
(left) Probability densities of recurrence times
in Southern California (SC) for the period 1984-2002,
for several $M_c$ values.
(right) The same probability densities rescaled by their rate. 
The data collapse                                                 \index{data collapse}
illustrates the fulfillment of a scaling law. 
The continuous line is a gamma fit.
}
\label{Dt_calif}
\end{figure}

We can illustrate this procedure with the
waveform cross-correlation catalog of Southern {California} 
obtained by Shearer et al.%
\footnote{Available at 
{\tt http://www.data.scec.org/ftp/catalogs/SHLK/}
} \cite{Shearer}
for the years 1984--2002, containing 26700 events
with $M\ge 2.5$ (84209 events with $M\ge 2$).
The recurrence-time probability densities for several 
values of $M_c$ are shown in Fig. \ref{Dt_calif}(left).
First, one can see that $\tau$ ranges from seconds to more
than 100 days
(in fact, we have restricted our analysis to recurrence-times
greater than one minute; shorter times do not follow the
same trend than the rest, probably due to the incompleteness
of the records in that time scale).
Also, the different distributions look very similar in shape,
although the ranges are different (obviously, the larger $M_c$,
the smaller the number of events $N(M_c)$,
and the larger the mean time between them).

Figure \ref{Dt_calif}(right) shows the same distributions  
but rescaled by the mean rate, as a function of the rescaled
recurrence time, i.e., $D(\tau;M_c)/R(M_c)$ versus $R(M_c)\tau$.
In this case all the distributions collapse onto a single curve $f$
and we can establish the fulfillment of a {\it scaling law}         \index{scaling law}
\cite{Corral_prl.2004},
\begin{equation}
    D(\tau;M_c)=R(M_c) f(R(M_c)\tau).
\label{scaling}
\end{equation}
where $f$ is the scaling function,                                  \index{scaling function}
and corresponds to the recurrence-time density
in the hypothetical case $R(M_c)=1$.
Note that we could have arrived to a similar equation
by scaling arguments, but there would be no reason 
for the function $f$ to be independent on $M_c$.
Only imposing the self-similarity of the process
in time-magnitude can lead to the fact that $f$ does not depend
on $M_c$ and therefore to the fact that $f$ is a scaling function.
As $R(M_c)$ verifies the GR law, the scaling law can be written
\begin{equation}
    D(\tau;M_c)=10^{-b M_c} \tilde f(10^{-b M_c} \tau).
\end{equation}
The GR law can be calculated from the scaling 
law; just calculate the mean recurrence time, 
$\langle \tau (M_c)\rangle = \int_0^\infty \tau 
D(\tau;M_c) d\tau = 10^{b M_c} \int_0^\infty z \tilde f(z)dz \propto 10^{b M_c}$, 
and as $\langle \tau (M_c)\rangle$ is the inverse of the mean rate,
then, $R(M_c) \propto 10^{-b M_c}$. 
But the scaling law does not
only include the GR law, it goes one step further, as it implies
that the GR law is fulfilled {\it at any time},
if times are properly selected;
indeed, events separated by recurrence times $\tau'$ for
$M\ge M_c'$ and $\tau$ for $M\ge M_c$
occur at a GR ratio, $10^{-b(M_c'-M_c)}$, if and only if the
ratio of the recurrence times is given by $10^{b(M_c'-M_c)}$.
Notice that the only requirement for the GR law 
to be fulfilled (for a long period of time)
is that $D(\tau;M_c)$ has a mean that verifies 
the GR law, i.e., $\langle \tau(M_c) \rangle =
R^{-1}(M_c) = R_0 10^{b M_c}$;
therefore, the fulfillment of the GR law at                 \index{instantaneous fulfillment of the Gutenberg-Richter law}
any time is a new feature of seismicity.

To make it more concrete, 
we can count the number of events in Southern California
with $M\ge 3$
coming after a recurrence time $\tau=100$ hours
and compare with the number of events with $M \ge 4$
after the same recurrence time; then the ratio of these 
numbers has nothing to do with the GR relation.
However, if for $M\ge 4$ we select events
with $\tau=1000$ hours (the $b-$value in
the GR law is very close to 1 in Southern California)
then, the number of these events is about 1/10 of 
the number of events with $M\ge 3$ and $\tau=100$ hours,
the same proportion as when we consider all events 
(no matter the value of $\tau$).
This could be somehow analogous to the well-known law of 
corresponding states in condensed-matter physics:
two pairs of consecutive earthquakes in different magnitude windows
would be in ``corresponding states''                         \index{law of corresponding states}
if their rescaled recurrence times are the same.

\subsection{Relation with the Omori Law}

In general, as seismicity is not stationary, the scaling function $f$
will change with the spatio-temporal window of observation.
In the case of Omori aftershock sequences,
the scaling function, and therefore the distribution 
of recurrence times, is related to the Omori law, as we now see.
Let us assume, just for simplicity, that the aftershock
sequence can be modeled as a nonhomogeneous Poisson process               \index{nonhomogeneous Poisson process}
(also called nonstationary Poisson process,                               \index{nonstationary Poisson process}
this is a Poisson process but with a time-variable rate,
in such a way that at any instant the probability of occurrence, 
per unit time, is not constant but is independent on the occurrence
of other events);
in this case the rate of occurrence will be 
given by the Omori law, Eq. (\ref{omori}).
Then, the recurrence-time density is a temporal mixture of Poisson
processes, which have a density $D(\tau | r(M_c)) = r e^{-r\tau}$, so,
\begin{equation}
D(\tau;M_c) = \frac{1}{\mu}\int_{r_m}^{r_0} r D(\tau | r) \rho(r;M_c) dr
= \frac{1}{\mu}\int_{r_m}^{r_0} r^2 e^{-r\tau} \rho(r;M_c) dr,
\label{mixing}
\end{equation}
where $\rho(r;M_c)$ is the density of rates, $\mu$ is a normalization
factor that turns out to be the mean value of $r$,
$\mu=\langle r(M_c) \rangle =\int r \rho(M_c) dr$
and $r_0(M_c)$ and $r_m(M_c)$ the maximum and minimum rate, respectively,
assuming $r_0\gg r_m$;
the factor $r$ appears because the probability of a given $D(\tau | r)$
to contribute to $D(\tau;M_c)$ is proportional to $r$.

The density of rates                                                 \index{density of seismic rates}
can be obtained by the projection of $r(t;M_c)$ onto the $r$ axis, 
turning out to be,
\begin{equation}
      \rho(r;M_c) \propto \left|\frac{dr}{dt}\right|^{-1}
\Rightarrow \rho(r;M_c) = \frac{C} {r^{1+1/p}}
\, \mbox{ for } \, r_m \le r \le r_0
\end{equation}
with $C$ just a constant (depending on $M_c$) that can be obtained from normalization.
Substituting, we get
\begin{equation}
D(\tau;M_c)=\frac C \mu \int_{r_m}^{r_0} r^{1-1/p} e^{-r\tau} dr
= \frac {C [\Gamma(2-1/p, r_m \tau)-\Gamma(2-1/p, r_0 \tau)]}
{\mu \tau^{2-1/p}},
\end{equation}
with $\Gamma(\alpha,z)\equiv \int_{z}^\infty z^{\alpha-1}e^{-z}dz $
the incomplete gamma function
(note that $\Gamma(1,z)=e^{-z}$).
It is clear that for intermediate recurrence times, 
$1/r_0 \ll \tau \ll 1/r_m$, we get a power law             \index{power-law distribution}        
of exponent $2-1/p$ for the recurrence time density, 
\begin{equation}
D(\tau;M_c) 
\simeq \frac {C \Gamma(2-1/p)} {\mu \tau^{2-1/p}},
\end{equation}
with $\Gamma(\alpha)$ the usual (complete) gamma function.
This power-law behavior has been derived before by Senshu
and by Utsu for nonhomogeneous Poisson processes
\cite{Utsu_handbook},
but our procedure can be easily extended beyond this case,
just defining a different $D(\tau|r)$,
for which the value of the recurrence-time exponent
$2-1/p$ is still valid.
Notice that the value of this exponent
is close to one if the $p-$value is close to one, 
but both exponents are only equal if $p=1$, in any other case
we have $2-1/p < p$, which means that in general $D(\tau;M_c)$
decays more slowly than $r(t;M_c)$.
If we consider large recurrence times, $r_m \tau \gg 1$,
we can use the asymptotic expansion 
$\Gamma(\alpha,z) \rightarrow z^{\alpha-1} e^{-z} + \cdots$
for $z\rightarrow \infty$ \cite{Abramowitz}, 
to get 
\begin{equation}
D(\tau;M_c) \simeq \frac{C r_m^{1-1/p}}{\mu} \frac{e^{-r_m\tau}}{\tau},
\label{Dtaularge}
\end{equation}
which in the limit we are working is essentially an exponential
decay.

Although the equations derived here for a nonhomogeneous 
Poisson process with Omori rate reproduce well
the recurrence-time distribution of aftershock sequences,
$D(\tau;M_c)$
\cite{Shcherbakov}, the choice of an exponential form for $D(\tau|r)$
is not justified, as we will see in the next sections.
Nevertheless, for the moment we are only interested in the form
of $D(\tau;M_c)$.

\subsection{Gamma Fit of the Scaling Function}

The fact that the density $D(\tau;M_c)$ 
for a nonhomogeneous Poisson-Omori sequence
is a power law for intermediate times
and follows Eq. (\ref{Dtaularge}) for long times suggests that a simpler
parameterization of the distribution can be obtained by
the combination of both behaviors;
in the case of the scaling function $f$,
which  must follow the same distribution as $D$
(but with mean equal to one), we can write
$$
f(\theta) \propto \frac{e^{-\theta/a}}{\theta^{2-1/p} (1+\theta)^{1/p \, -1}}.
$$
However, as both power laws of $\theta$ are very similar
and in the long time limit it is the exponential alone
what is really important, we can simplify even further
and use the gamma distribution                                     \index{gamma distribution}
to model $f$; so,
\begin{equation}
    f(\theta) = \frac C {a \Gamma(\gamma)} 
\left(\frac a {\theta}\right)^{1-\gamma} e^{-\theta/a},
\label{gamma}
\end{equation}
where $\theta$ plays the role of a dimensionless recurrence time,       \index{dimensionless recurrence time}
$\theta \equiv R\tau$, $a$ is a dimensionless scale parameter, and 
$C$ is a correction to normalization due to the fact
that the gamma distribution may not be valid for very short times;
this will allow the shape parameter $\gamma$
not to be restricted to the case $\gamma > 0$,
the usual condition for the gamma distribution
(nevertheless, if $\gamma \le 0$ the factor
$\Gamma(\gamma)$ is inappropriate for normalization).
As $f$ is introduced in such a way that the mean of $ \theta$
is $\langle \theta \rangle= 1$, the parameters are not independent;
for instance, for $C=1$,  $\langle \theta \rangle= \gamma a$
and in consequence $a=1/\gamma$.  
So, essentially, we only have one parameter to fit, $\gamma$,
to characterize the process.
In the case of Omori sequences, 
$1-\gamma=2-1/p$ and $a=R/r_m$, $\Rightarrow \gamma \simeq r_m /R$,
but we will see that the gamma distribution
has a wider applicability than just Omori sequences.

A fit of the gamma distribution to the rescaled distribution 
for Southern-California, shown in Fig. \ref{Dt_calif}(right),
yields the parameter values $\gamma \simeq 0.22$ and $a \simeq 3$;
this yields a power-law                                              \index{power-law distribution}
exponent for small and intermediate times
$1-\gamma \simeq 0.78$ and
allows to calculate a $p-$value $p= (1+\gamma)^{-1} \simeq  0.82$,
which can be interpreted as an average for Southern California, and 
a minimum rate $r_m\simeq R/3$.
Of course, with our resolution we only can establish 
$1-\gamma \simeq p \simeq 0.8$.

\subsection{Universal Scaling Law for Stationary Seismicity}

We have mentioned the nonstationary character of seismicity
and that in consequence the scaling function $f$ depends on
the window of observation.
A more robust, universal law can be established  
if we restrict our study to stationary seismicity.                         \index{stationary seismicity}
By stationary seismicity we mean in fact homogeneity in time,
which implies that the statistical properties of the process do not depend
on the time window of observation, in particular, the mean rate
must be practically constant in time. 

It is obvious that an aftershock 
sequence following the Omori law (with $p >0$) is not stationary,
but observational evidence shows that in other cases seismicity
can be well described by a stationary process, for example worldwide
seismicity for the last 30 years (for which there are reasonably good data)
or regional seismicity in between large aftershock sequences.
It should be clear that considering stationary seismicity has nothing to
do with declustering (the removal of aftershocks from data).
We simply consider periods of time for which no aftershock sequence 
dominates in the spatial region selected for study,
but many smaller sequences may be hidden in the data, 
intertwined in such a way to give rise to an
overall stationary seismic rate.

\begin{figure}
\centering
\includegraphics[height=6cm]{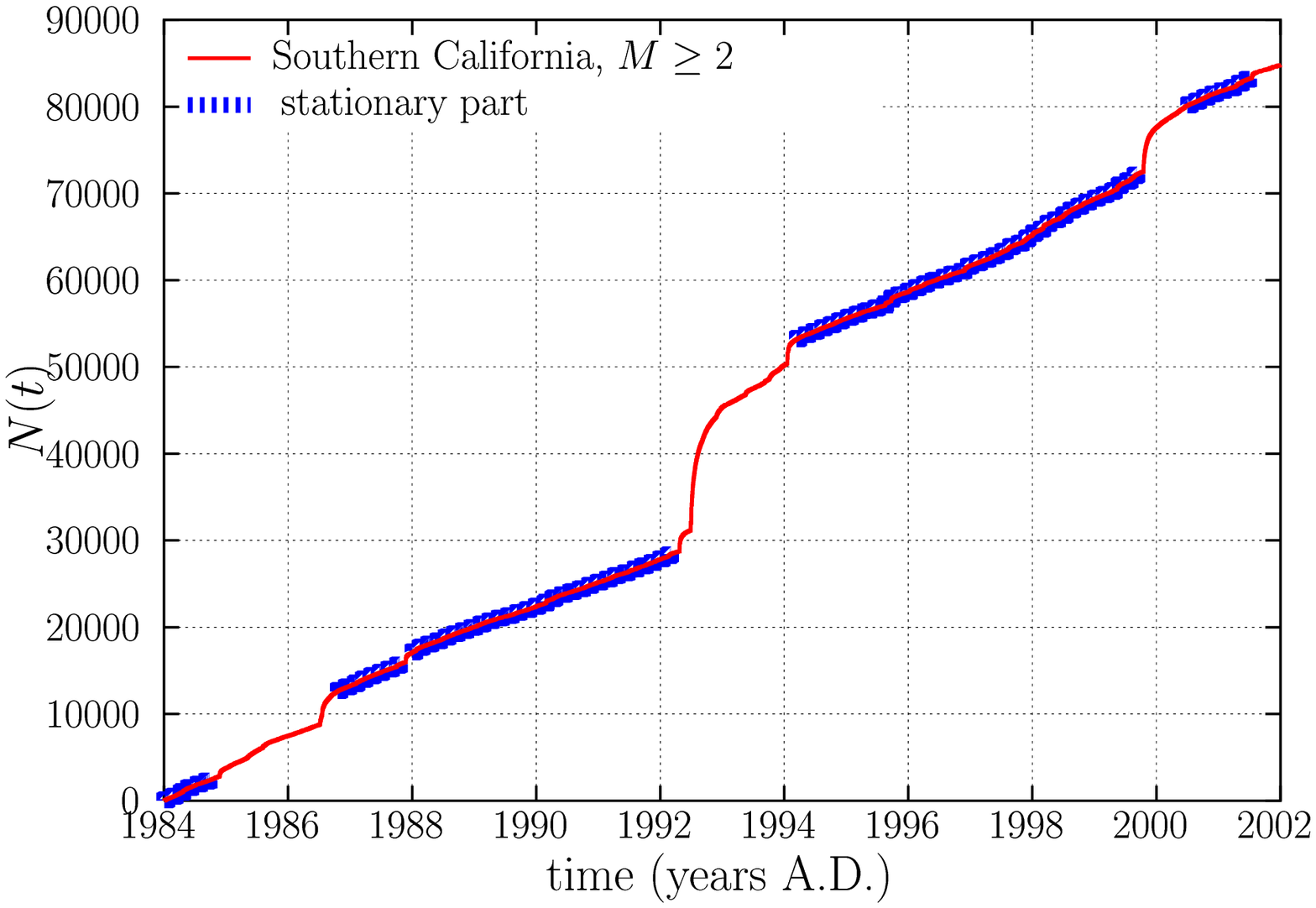}
\caption{
Accumulated number of earthquakes in Southern
California as a function of time. 
Some stationary or nearly stationary periods mentioned in the text
are specially marked, see subsection \ref{mm}.
}
\label{Nt}       
\end{figure}

The total number of earthquakes in Southern-California
(from Shearer et al.'s catalog)
as a function of time since 1984 
is displayed in Fig. \ref{Nt}.
Clearly, the behavior of the number of earthquakes
in time is nonlinear, with episodic abrupt increments which
correspond to large aftershock sequences, following
the trend prescribed by the Omori law, 
$N(M_c,t) = N(M_c,0)  + \int_0^t r_0(M_c)/(1+t'/c)^p dt'$.
However, there exist some periods which follow a linear increase
of $N(M_c,t)$ versus $t$; in particular, we have chosen for analysis
the intervals (in years, with decimal notation) 
1984--1986.5,
1990.3--1992.1,
1994.6--1995.6,
1996.1--1996.5,
1997--1997.6,
1997.75--1998.15,
1998.25--1999.35,
2000.55--2000.8,
2000.9--2001.25,
2001.6--2002,
and
2002.5--2003.
These intervals comprise a total time span of 9.25 years 
and contain 6072 events for $M\ge 2.5$,
corresponding to a mean rate $R(2.5)=1.7 $ earthquakes/day.
Note from the figure that not only the rate of occurrence is
nearly constant for each interval, but different intervals have 
similar values of the rate.

\begin{figure}
\centering
\includegraphics[height=4.1cm]{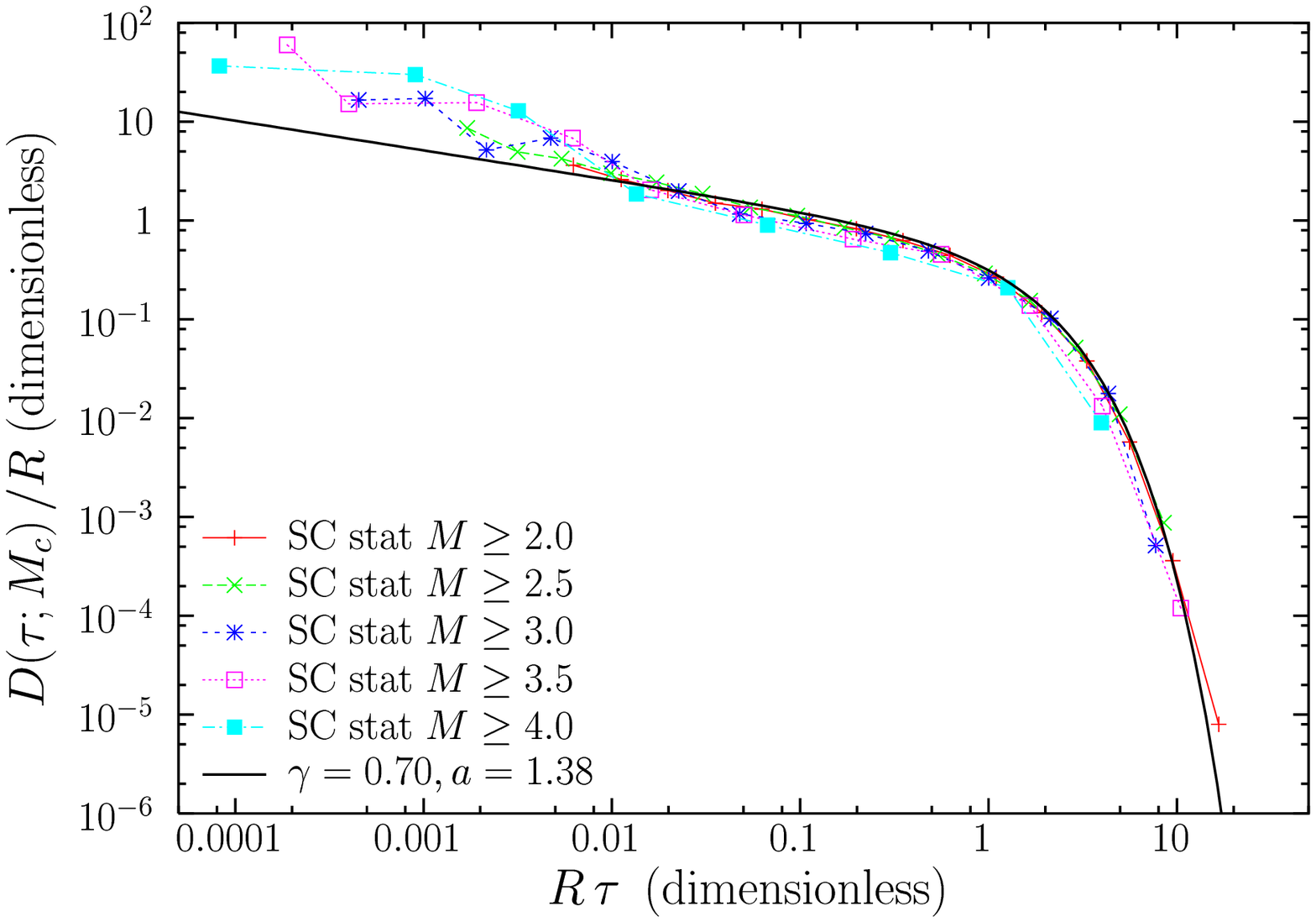}
\includegraphics[height=4.1cm]{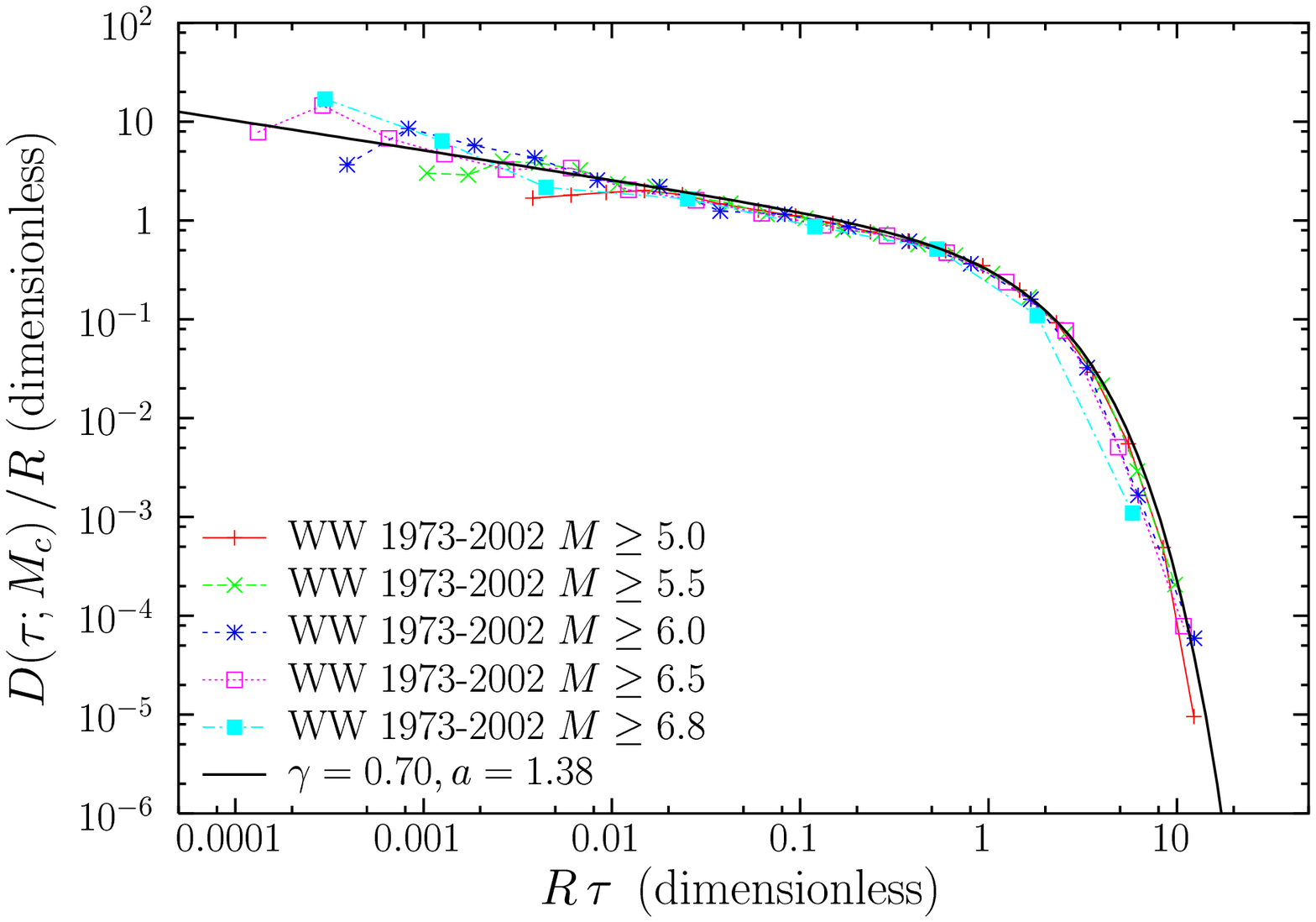}
\caption{
Rescaled
recurrence-time probability densities for the stationary periods
explained in the text for Southern California (left) and for 
worldwide seismicity (right).
The solid line is the same function in both cases,
showing the universal character of the scaling law fulfilled.
}
\label{Dtsca2}
\end{figure}

We will study all these stationary periods together, 
in order to improve the statistics.
The probability densities of the recurrence times are calculated
from all the periods 
and the corresponding rescaled distributions appear in Fig. \ref{Dtsca2}(left).
The good quality of the data collapse indicates the validity 
of a scaling law of the type of  Eq. (\ref{scaling}), although the scaling function
$f$ is clearly different than the one for 
the whole time period 1984-2002, in particular, 
the power-law is much flatter, which is an indication
that the clustering degree is smaller in this case, 
in comparison, but still exists. 
Nevertheless, it is remarkable that this kind of clustering is different
than the clustering of aftershock sequences, as
in this case we are dealing with a stationary process.
The figure shows also a plot of the scaling function $f$
parameterized with a gamma distribution with $\gamma=0.7$ and  $a=1.38$,
which indeed implies a power-law exponent $1-\gamma=0.3$.

We now present the results for recurrence times
in worldwide scale, using the 
NEIC-PDE worldwide catalog
(National Earthquake Information Center, Preliminary
Determination of Epicenters
\footnote{Available at
{\tt http://wwwneic.cr.usgs.gov/neis/epic/epic$\_$global.html}
}) 
which covers the period 1973-2002 and yields 46055 events
with $M\ge 5$. In this case the total number of earthquakes
grows linearly in time, which confirms the stationarity of
worldwide seismicity.
The corresponding rescaled recurrence-time 
probability densities are shown in Fig. \ref{Dtsca2}(right),
together with the scaling function used in the previous case
(i.e., Southern-California stationary seismicity). The collapse of
the data onto a single curve is again an indication of the 
validity of a scaling law, and the fact that this curve
is well fit by the same scaling function than in the Southern-California
stationary case is a sign of {\it universality}.                             \index{universality}
We use the term universality with the usual meaning in statistical
physics, in which it refers to very different systems
(gases or magnetic solids, or in our case seismic occurrence
in quite diverse tectonic environments) sharing the same quantitative
properties.

In fact, the universality of the scaling law for recurrence-time
distributions in the stationary case has been tested 
for several other regions, namely, Japan, Spain, New Zealand, New Madrid
(USA), and Great Britain, with magnitude values ranging
from $M\ge 1.5$ to $M\ge 7.5$ (which is equivalent
to a factor $10^9$ in the minimum dissipated energy),
and for spatial areas as small as $0.16^\circ \simeq 20$ km
\cite{Corral_prl.2004,Corral_pre.2005}

\subsection{Universal Scaling Law for Omori Sequences}

We now return to nonstationary seismicity to show how the 
universal scaling law for recurrence times applies there.
For this purpose, let us consider the Landers earthquake,                 \index{Landers earthquake} 
with magnitude $M=7.3$, 
the largest event in Southern California in the last decades,
taking place in 1992, June 28, at $34.12^\circ N$, $116.26^\circ W$.
After the earthquake, seismicity in Southern California
followed the usual behavior when large shallow events happen:
a sudden enormous increase in the number of earthquakes and a
consequent slow decay in time,
in good agreement with the Omori law.

The previous universal results for stationary seismicity
can be generalized in the nonstationary case by replacing
the mean seismic rate $R(M_c)$ by the ``instantaneous''
seismic rate $r(t,M_c)$ as the scaling factor in Eq. (\ref{scaling}). 
Then, in order to obtain the rescaled, dimensionless recurrence time       \index{dimensionless recurrence time}
$\theta$,
we will rescale each recurrence time as
\begin{equation}
\theta_i \equiv r(t_i;M_c) \tau_i.
\end{equation}
This means that it is the instantaneous rate of occurrence
which sets the time scale.

\begin{figure} 
\centering
\includegraphics[height=4.1cm]{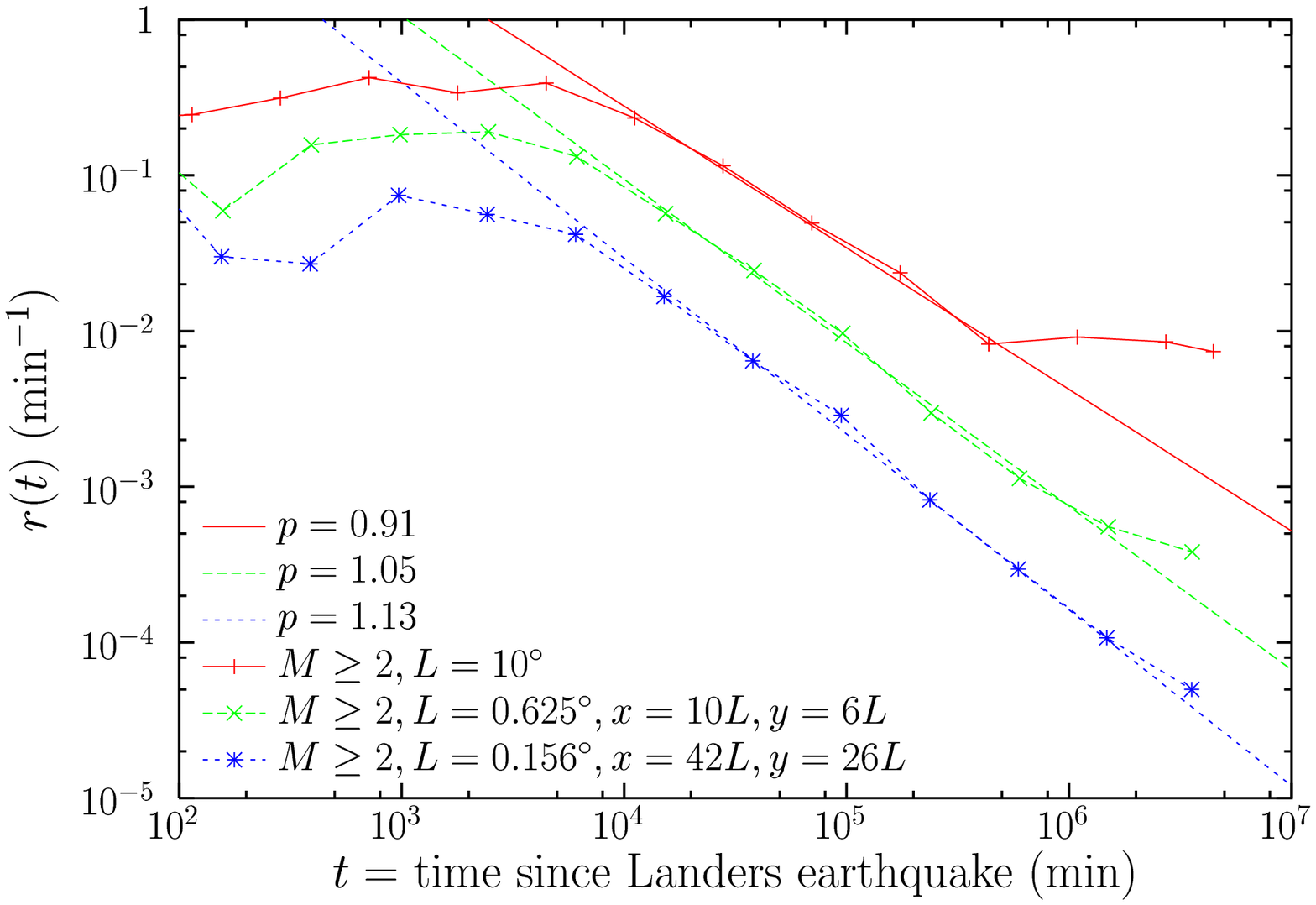}
\includegraphics[height=4.1cm]{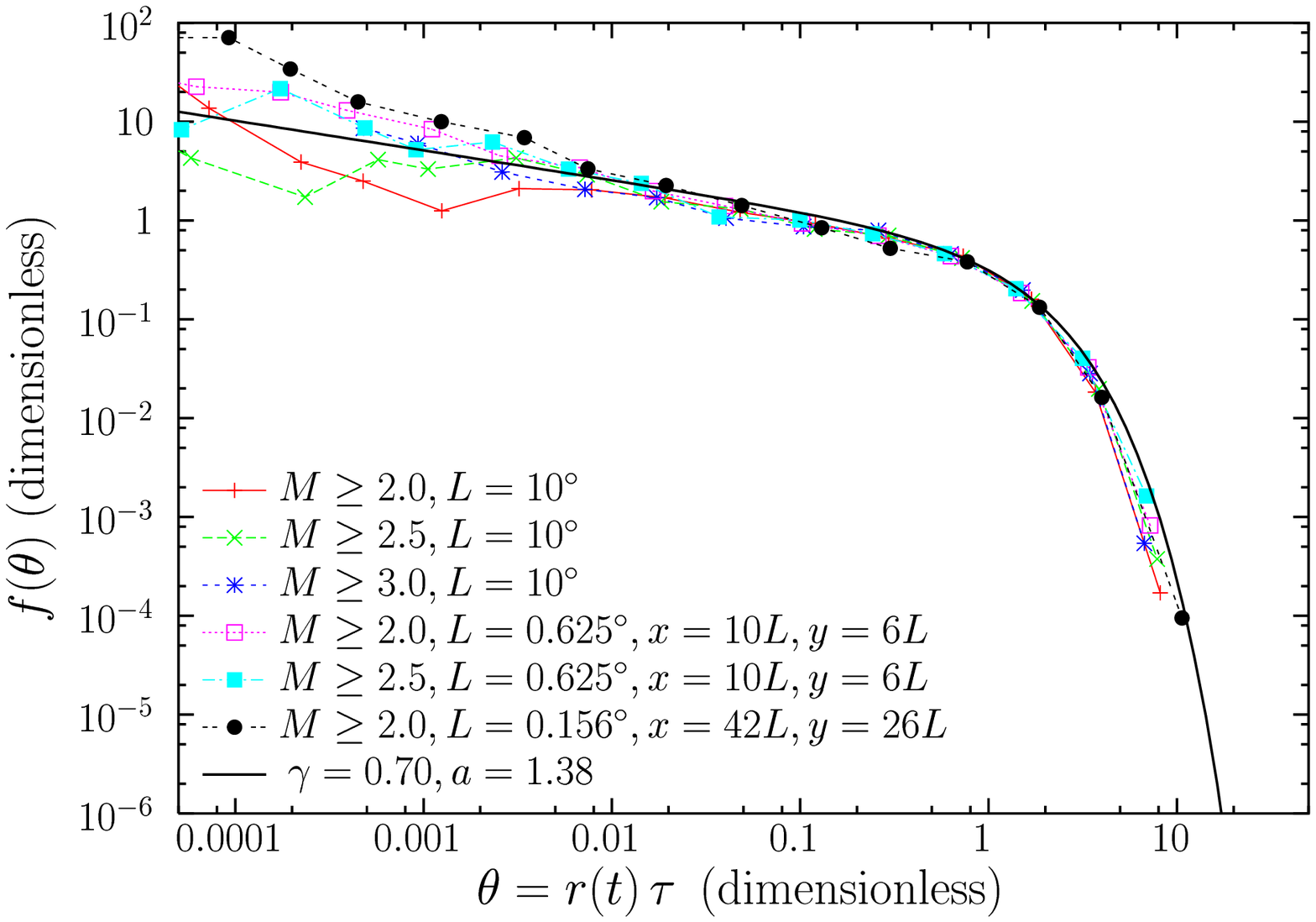}
\caption{
(left)
Seismic rate as a function of the time elapsed since the Landers earthquake
for regions of different size $L$ including the event,
using the SCSN catalog.
Only events with $M \ge 2$ are considered.
The straight lines correspond to power laws,
with exponents given by the $p-$value.
(right)
Recurrence-time probability densities for the power-law
regime of the decay of the rate after the Landers event,
rescaled at each time by the rate.
The solid line represents the universal scaling function 
in terms of a gamma distribution.
\label{rt_Landers}
}
\end{figure}

First, we examine the complete seismicity with $M \ge M_c$
for a square region
(in a space in which longitude and latitude are considered as rectangular
coordinates), the region containing the Landers event (but not centered 
on it);
that is, as in previous sections, we will not
separate aftershocks from the rest of events.
As expected, after a few days from the mainshock, 
the seismic rate decreases as a pure power law,
which is equivalent to take 
$t \gg c$ in Eq. (\ref{omori}), so,
$$
r(t;M_c) = r_0 \left(\frac c t \right)^p,
$$
which lasts until the rate reaches the background seismic level.
This form for $r(t;M_c)$ is fit to the measured seismic rate,
see Fig. \ref{rt_Landers}(right). 
One advantage of analyzing the pure power-law regime only, 
rather than the whole sequence using the modified Omori law, 
Eq. (\ref{omori}), is that it is believed that the
deviations from power-law behavior for short
times are due to the incompleteness of the catalogs
after strong events; therefore, in this way we avoid the problem
of incompleteness.

Next, 
using the results of the fit rather than
the direct measurement of $r(t;M_c)$
we calculate
\begin{equation}
\theta_i = 
r_0 \tau_i\left(\frac c {t_i} \right)^p.
\end{equation}
In fact, this rescaling could be replaced by 
$\theta_i = r(t_{i-1}) \tau_i$
or by $\theta_i = r(t_{i-1}+\tau_i/2) \tau_i$,
with no noticeable difference in the results,
as the rate varies 
very slowly at the scale of the
recurrence time.

The probability densities of the rescaled recurrence times $\theta_i$
obtained in this way
are displayed in Fig. \ref{rt_Landers}(right),
showing a slow power-law decay followed
by a faster decay,
in surprising agreement, not only qualitative but also
quantitative, with the results
for stationary seismicity,
in such a way that the universal scaling function
for the stationary case is still valid 
\cite{Corral_prl.2004,Corral_npg.2005}. 
Therefore, as in that case, 
the power-law regime in the density 
is a sign of clustering, but as the 
primary clustering structure of the sequence
has been removed by the rescaling with Omori rate, this
implies the existence of a secondary clustering structure            \index{clustering structure, secondary}
inside the main sequence, due to the fact that any large
aftershock may generate its own aftershocks 
\cite{Ogata99,Helmstetter_pre02}.
What is remarkable is that this structure seems to be identical
to the one corresponding to stationary seismicity.

An important consequence of this is that the time behavior
of seismicity depends on just one variable:
the seismic rate.
Another implication is the fact that
aftershock sequences cannot be described as a nonhomogeneous Poisson process,
as in that case one should obtain an exponential distribution for $f(\theta)$.
The use of the nonhomegeneous Poisson process previously in this work 
must be understood only as a first approximation to justify 
the use of the gamma fit for 
the distribution of recurrence times
in an Omori sequence.
Nevertheless, we will see in the next sections that the generalization of
the nonhomogeneous Poisson process taking into account the results
explained here
leads to similar conclusions for the time distribution
in the sequence.

The rescaling of the recurrence times with the seismic rate
$r(t,M_c)$ can be applied also to the occurrence times $t_i$,
in order to transform the Omori sequence (or in general 
any sequence with a time-variable rate) into a stationary sequence.
For this purpose we define the
accumulated rescaled recurrence time $\Theta$,
defined as $\Theta_i=\theta_1+\theta_2+\cdots+ \theta_i$,
which plays the role of a stationary occurrence time, 
in the same way that $\theta_i$ plays the role of
a stationary recurrence time, in general.
This allows the complete comparison between stationary
seismicity and aftershock sequences.

\section{The Paradox of the Decreasing Hazard Rate and
the Increasing Time Until the Next Earthquake}

Other functions, in addition to the probability density,
are suitable for describing the general properties of
recurrence times. Although from a mathematical point of view
the functions we are going to introduce are fully equivalent
to the probability density, they show much clearly
some interesting temporal features of seismicity.

\subsection{Decreasing of the Hazard Rate }

Let us consider first the {\it hazard rate},                 \index{hazard rate}
$\lambda(\tau;M_c)$,
defined for a certain region and for $M \ge M_c$ as
the probability per unit time of 
an immediate earthquake given that there has been a period 
$\tau$ without activity \cite{Kalbfleisch},
$$
\lambda(\tau;M_c) \equiv 
       \frac{ \mbox{Prob} [\tau < \tau' \le \tau + 
       d \tau \, | \, \tau' > \tau ]}{d \tau}
       =\frac{D(\tau;M_c)}{S(\tau;M_c)},
$$
where 
$\tau'$ is a generic label for the recurrence time,
while $\tau $ refers to a particular value of the same quantity,
the symbol $|$ denotes conditional probability,
and $S(\tau;M_c)$ is the survivor function, 
$S(\tau;M_c)\equiv \mbox{Prob}[\tau' > \tau] = \int_\tau^\infty D(\tau';M_c)d\tau'$.
Introducing the scaling law (\ref{scaling}) for $D$
in the definitions
it is immediate to obtain that both $S(\tau;M_c)$ and $\lambda(\tau;M_c)$ 
verify also scaling relations, $S(\tau;M_c)=g(R\tau)$ and $\lambda(\tau;M_c)=R h(R \tau)$.
If we make use of the gamma parameterization, Eq. (\ref{gamma}),
with $C\simeq 1$,
we get for the scaling function $h$,
$$
h(\theta)=\frac{1}{a }  \left(\frac{a}{\theta} \right)^{1-\gamma}
           \frac{e^{-\theta/a}}{\Gamma(\gamma,\theta/a)}.
$$
For short recurrence times this function
diverges as a power law,
$$
     h(\theta) \simeq \frac{1}{\Gamma(\gamma)a^\gamma \theta^{1-\gamma}}
$$
(in fact, in this limit the hazard rate becomes undistinguishable
from the probability density).
On the other hand, $h(\theta)$
tends as a power law to the value $1/a$ as $\theta \rightarrow \infty$;
indeed, making use of the expansion 
$\Gamma(\gamma,z) \rightarrow z^{\gamma-1} e^{-z}
[1-(1-\gamma)/z + (1-\gamma)(2-\gamma)/z^2+\cdots]$
\cite{Abramowitz}, we get
$$
    h(\theta) = \frac 1 a \left[1 +\frac{a(1-\gamma)}{\theta} +\cdots\right].
$$
The overall behavior for $\gamma < 1$ is that $h(\theta)$ decreases monotonically
as $\theta$ increases;
so, contrary to common belief and certainly counterintuitively, 
these calculations allow us to predict that
the hazard does not increase with
the elapsed time since the last earthquake, but just the opposite,
it decreases up to an asymptotic value that corresponds to
a Poisson process of rate $R/a$.
This means that although the hazard rate decreases, 
it never reaches the zero value, and sooner or later
a new earthquake will strike.

If we compare the hazard rate for $\gamma < 1$ with 
that of a Poisson process with the same mean
(given by $\gamma=a=1$, and rate $R$), we see that for short 
recurrence times
the hazard rate is well above the Poisson value,
implying that at any instant the probability 
of having an earthquake is higher than in the Poisson case.
In contrast, for long times the probability is below
the Poisson value, by a factor $1/a$.
This is precisely the most direct characterization 
of clustering in time, for which we can say that events tend to 
attract each other,
being closer in short time scales and more separated
in long time scales (in comparison with the Poisson process).

In conclusion, we predict that seismicity is clustered 
independently on the scale of observation;
in the case of stationary seismicity
this clustering is much less trivial than the clustering due 
to the increasing of the rate in aftershock sequences.
Taking advantage of the self-similarity implied by the scaling law,
we could extrapolate the clustering behavior to the largest
events worldwide ($M\ge 7.5$) over relatively small spatial 
scales (hundreds of kilometers) and we would obtain a behavior
akin to the long-term clustering observed by other means
\cite{Kagan_91}

\subsection{Increasing of the Residual Time Until the Next Earthquake}

Let us introduce now the {\it expected residual recurrence time},     \index{expected residual recurrence time}
$\epsilon(\tau_0;M_c)$,
which provides the expected time till the next earthquake,
given that a period $\tau_0$ without earthquakes 
(in the spatial area and range of magnitudes considered) has elapsed
\cite{Kalbfleisch},
$$
\epsilon(\tau_0;M_c) \equiv
\langle \tau-\tau_0 \, | \, \tau > \tau_0 \rangle =
\frac{1}{S(\tau_0;M_c)}
\int^{\infty}_{\tau_0} (\tau-\tau_0) D(\tau;M_c) d \tau.
$$
where $|$ denotes that the mean is calculated only
when the condition $\tau > \tau_0$ is fulfilled.
Again, the scaling law for $D$ implies a scaling form for this function, 
which is $\epsilon(\tau_0)=e(R \tau_0) / R$, and introducing the gamma
parameterization we get for the scaling function 
$$
e(\theta)
=a \frac{\Gamma({\gamma+1},\theta/a)}{\Gamma({\gamma},\theta/a)} - \theta
=a \left[ \gamma + \left(\frac \theta a\right)^\gamma 
\frac{e^{-\theta/a}}{\Gamma(\gamma,\theta/a)} \right] -\theta,
$$
making use of the relation 
$\Gamma(\gamma+1,z)=\gamma\Gamma(\gamma,z) + z^\gamma e^{-z}$.
For short times we obtain 
$$
e(\theta) = a\gamma +\frac{a}{\Gamma(\gamma)} 
\left(\frac\theta a\right)^\gamma -\theta+\cdots;
$$ 
remember that the unconditional mean is 
$\langle \theta \rangle = \gamma a=1$,
precisely the value obtained for $\theta=0$.
For long times $e(\theta)$ reaches, 
again as a power law, an asymptotic value equal to $a$, i.e.,
$$
e(\theta) 
= a \left[1 -\frac{a(1-\gamma)}{\theta} +\cdots\right].
$$
The global behavior of $e(\theta)$ is monotonically 
increasing as a function of $\theta$ if $\gamma < 1$.
Therefore, the residual time until the next earthquake should grow with the elapsed
time since the last one.
Notice the counterintuitive behavior that this represents:
if we decompose the recurrence time $\tau$ as $\tau=\tau_0 + \tau_f$,
with $\tau_f$ the residual time to the next event, the increase of $\tau_0$
implies the increase of the mean value of $\tau_f$, but the mean value
of $\tau$ is kept fixed.
In fact, 
this is fully equivalent to the
previously reported decreasing-hazard phenomenon
and
just a more dramatic version of the classical 
waiting-time paradox                                          \index{waiting-time paradox}
\cite{Feller,Szekely,Schroeder}.

This result seems indeed paradoxical for any time process,
as we naturally expect that the residual recurrence (or waiting)
time decreases as time increases; think for instance that you are
waiting for the subway: you are confident that the next train
is approaching; or when you celebrate your birthday,
your expected residual lifetime decreases (at any time, in fact).
Of course, for an expert statistician the case of earthquakes 
is not paradoxical, but only counterintuitive, 
and he or she can provide the counterexamples 
of newborns (mainly in underdeveloped countries) or of private 
companies, which become healthier or more solid as time passes 
and therefore their expected residual lifetime increases with time.
These counterintuitive behaviors can be referred to 
as a phenomenon of {\it negative aging}.                              \index{negative aging}

Nevertheless, for the concrete case of earthquakes 
the increasing of the expected residual recurrence time
is still
paradoxical,
since one naively expects that the longer the time one has been waiting
for an earthquake, the closer it will be,
due to the fact that as time passes 
stress increases on the faults
and the next earthquake becomes more likely.
Nevertheless, note that our approach does not deal
with individual faults but with two-dimensional,
extended regions, and in this case the evolution of
the stress is not so clear.
It is worth mentioning that, as far as the author knows,
no conclusive study of this kind has been performed for observational data
in individual faults, the difficulty on associating earthquakes to faults
is one the major problems here.

\subsection{Direct Empirical Evidence}

Our predictions for earthquake recurrence times follow
the line initiated by other authors.
Davis {et al.} \cite{Davis},
pointed out that when a lognormal distribution is a priori assumed
for the recurrence times, the expected residual time increases
with the elapsed time.
However, the increase there was associated to the update of the
distribution parameters as the time since the last earthquake
(which was taken into account in the estimation) increased,
and not to an intrinsic property of the distribution.
Sornette and Knopoff \cite{Sornette_Knopoff} showed that 
the increase (or decrease) depends completely
on the election of the distribution, and studied 
the properties of a number of them.
We now will see that the observational data provide 
direct and clear evidence in favor of the picture of 
an incoming earthquake which is moving away in time.

\begin{figure}
\centering
\includegraphics[height=4.1cm]{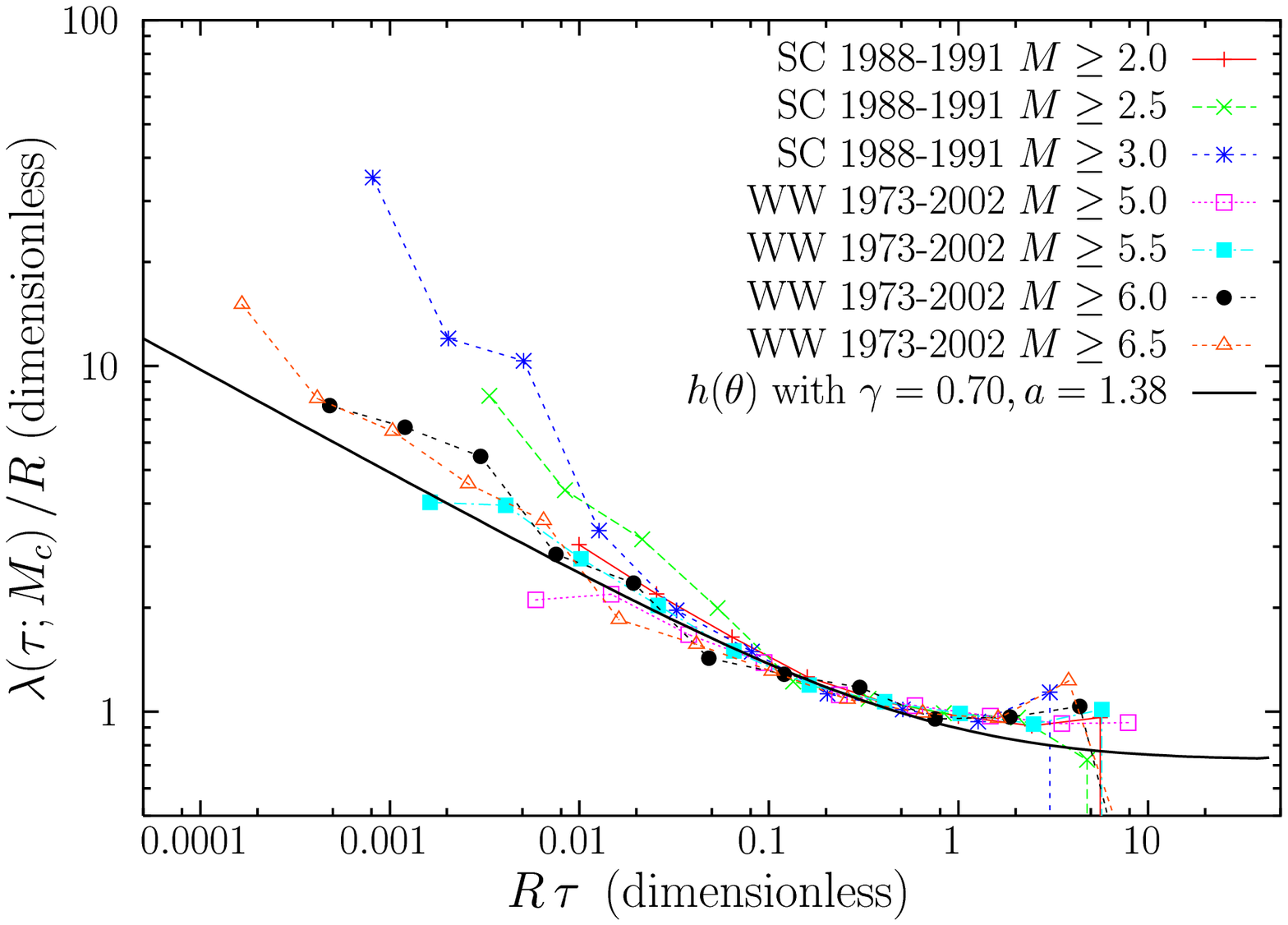}
\includegraphics[height=4.1cm]{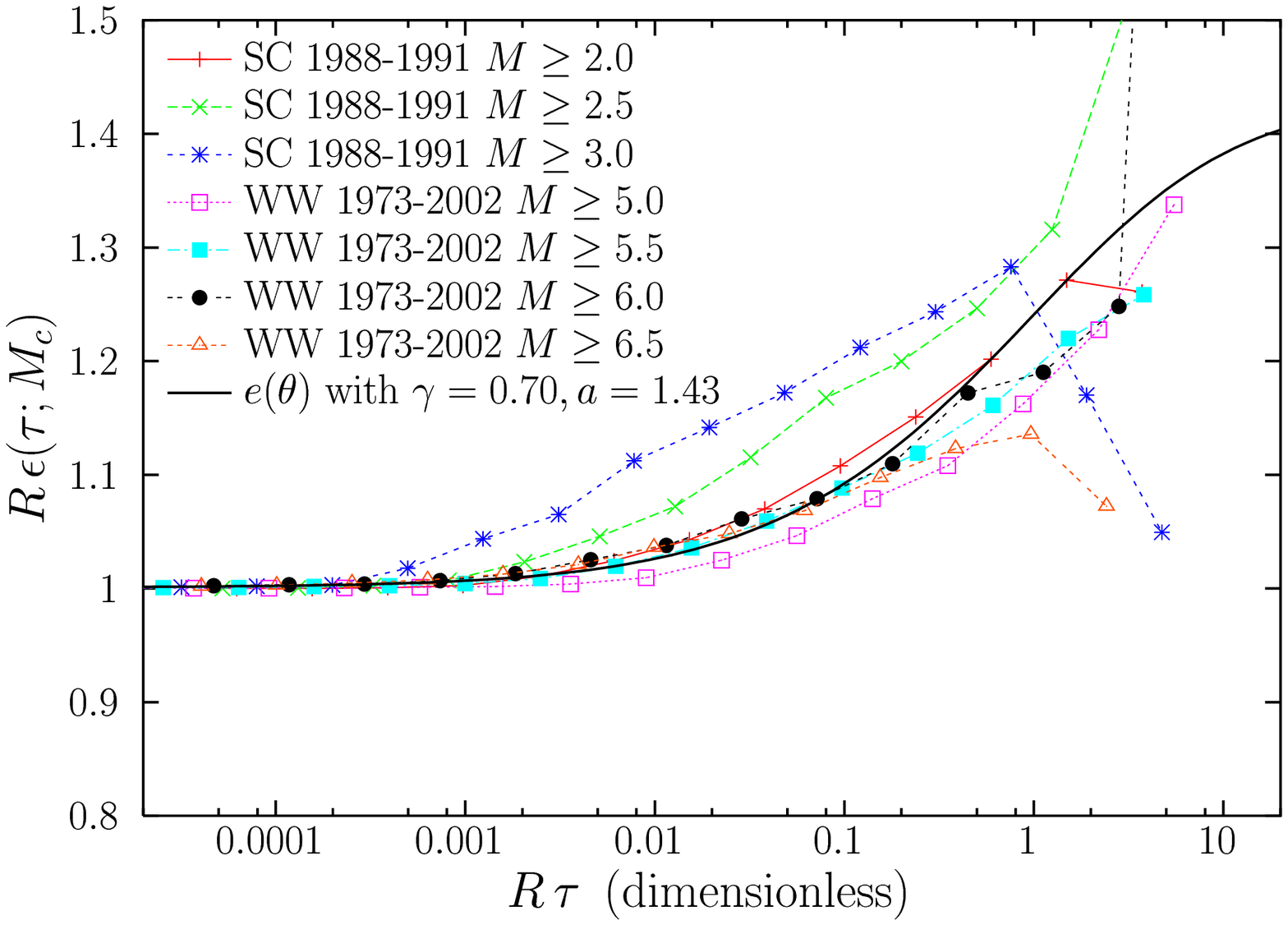}
\caption{Rescaled hazard rate (left)
and rescaled expected residual recurrence time
(right) as a function of time
for Southern California, 1988-1991 
(nearly stationary period) and for worldwide
seismicity, 1973-2002.
The observational data agrees with the scaling functions
derived from the gamma distribution.
In the right plot the parameter $a$ is not free,
but $a=1/\gamma$ to enforce $e(0)=1$.}
\label{hazard}
\end{figure}

Indeed,
in order to rule out the possibility that these paradoxical
predictions are an artifact introduced by the gamma parameterization,
we must contrast them with real seismicity; in fact,
both the hazard rate and the expected 
residual recurrence time
can be directly measured from the catalogs, 
with no assumption about their functional form.
Their definitions provide a simple way to estimate
these functions, and in this way we have applied 
these definitions to the recurrence-time data
\cite{Corral_pre.2005}.
From the results displayed in Fig. \ref{hazard}
it is apparent that in all cases 
the hazard rate decreases with time whereas the expected
residual recurrence time increases, as we have predicted.
Although both quantities are well approximated by the proposed universal
scaling functions, 
we emphasize that their behavior does not depend on any modeling
of the process and in particular is independent on the gamma
parameterization.
Moreover, the fact that $\epsilon(\tau_0)$ is far from being constant at large times
means that the time evolution is not properly described by a Poisson process,
even in the long-time limit.

We conclude stating that the contents of this section can be summarized
in this simple sentence: 
{\it the longer since the last earthquake, 
the lowest the hazard for a new one,
}
which is fully equivalent to this one (although less shocking):
{\it the longer since the last earthquake, 
the longer the expected time till the next.
}
Moreover, this happens in a self-similar way,
thanks to the scaling laws which are fulfilled.

\section{Scaling Law Fulfillment as Invariance
Under a Renormalization-Group Transformation}

It is interesting to realize that the scaling law for the recurrence-time
distribution, Eq. (\ref{scaling}), implies the invariance of the distribution under a 
renormalization-group transformation.
Let us investigate deeper the meaning of the scaling analysis
we have performed and its relation with the renormalization group.

\begin{figure}
\centering
\includegraphics[height=3cm]{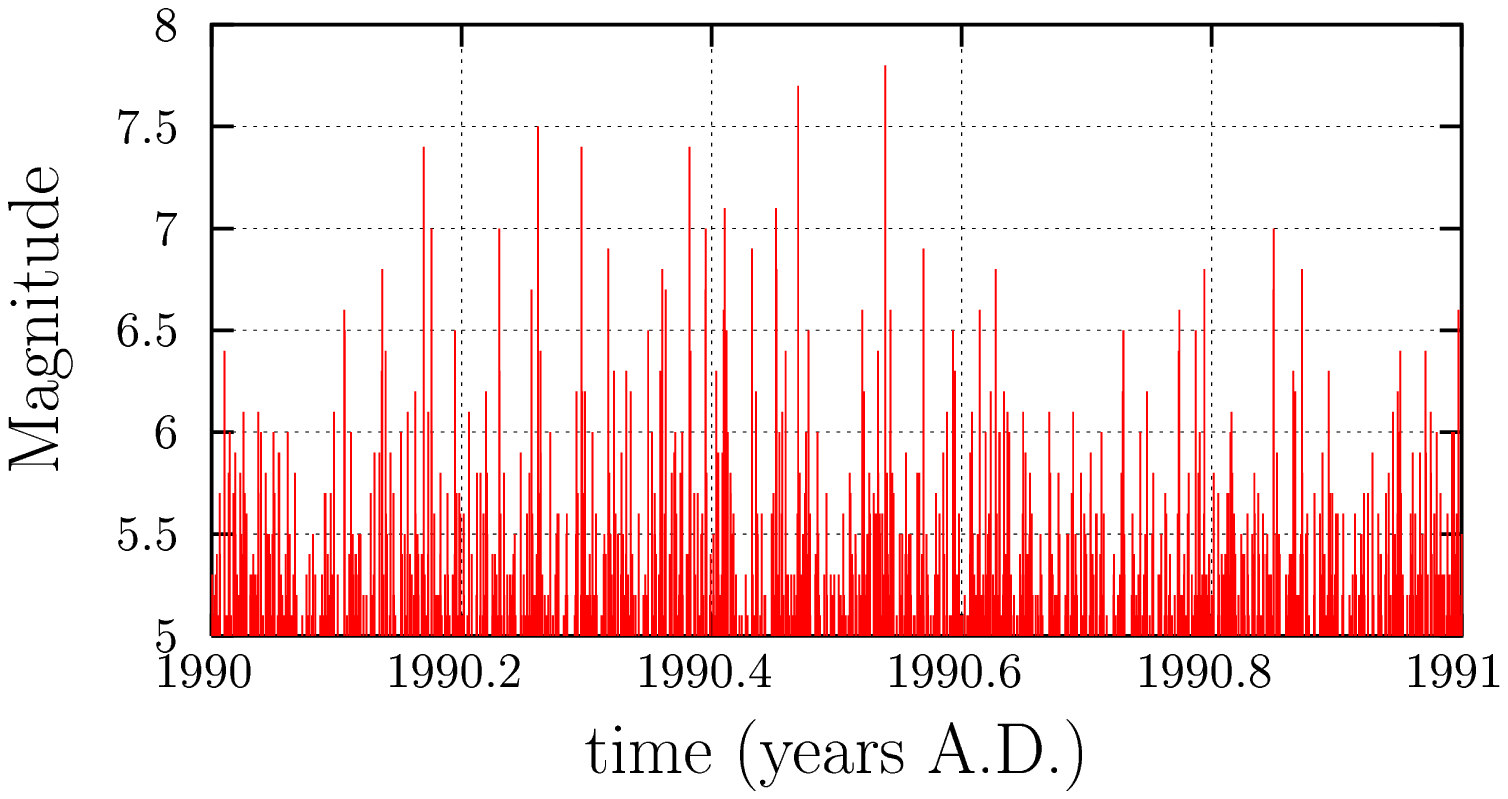}
\includegraphics[height=3cm]{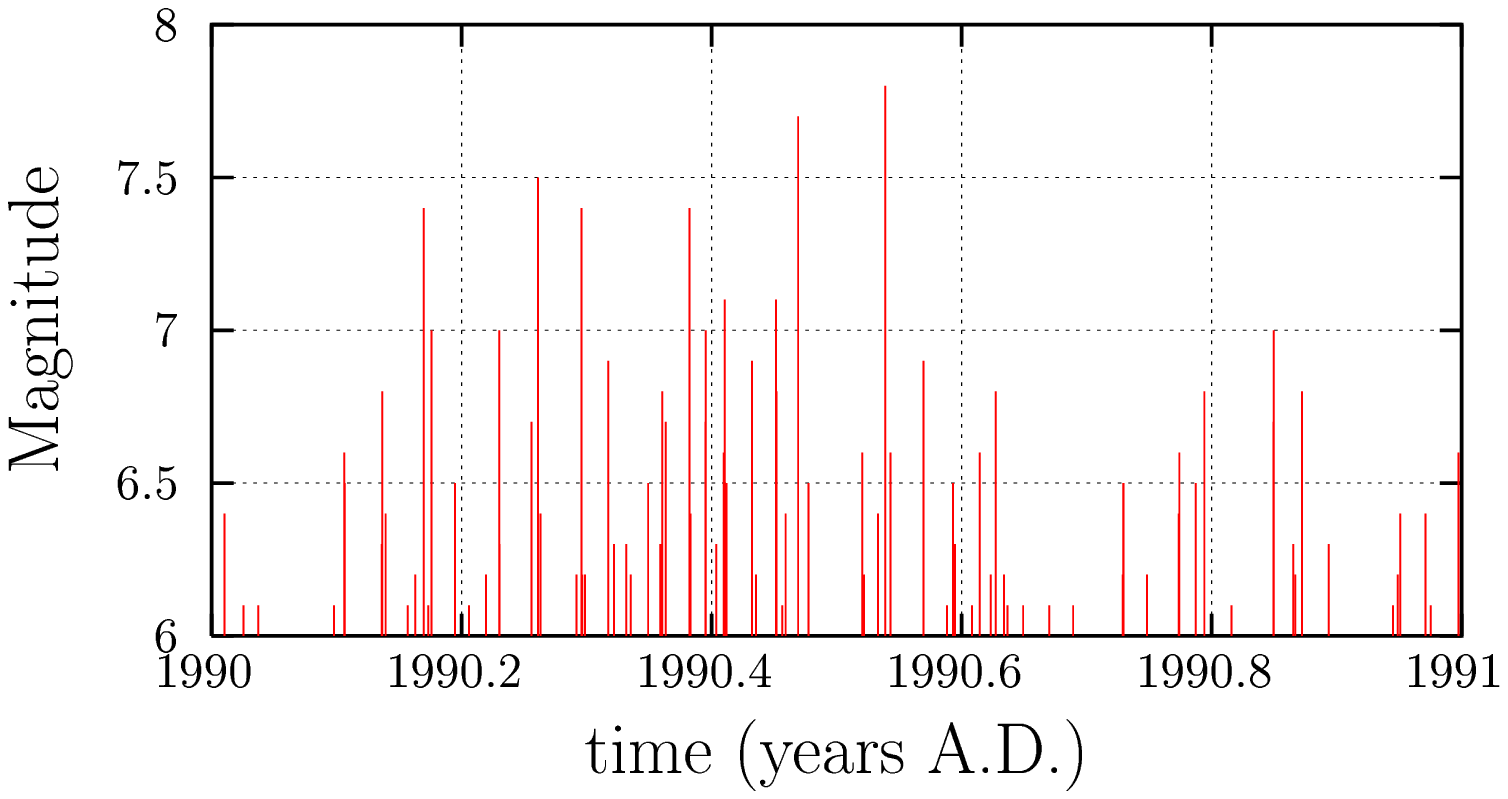}
\includegraphics[height=3cm]{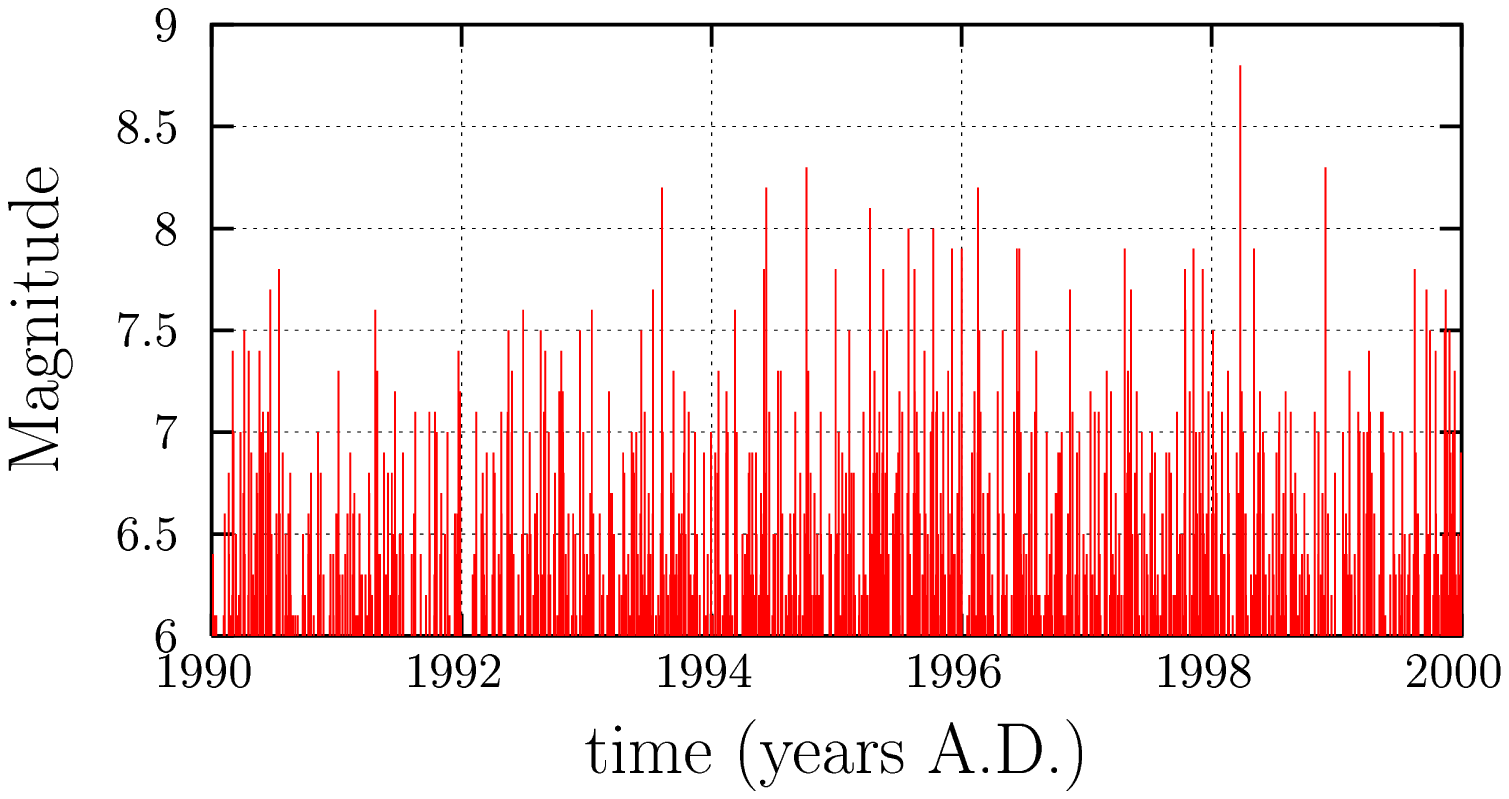}
\caption{
Magnitude versus time of occurrence of worldwide 
earthquakes for several magnitude-time windows.
The rising of the magnitude threshold from 5 to 6
illustrates the thinning or decimation process characteristic
of the first step of a renormalization-group transformation.
The second step is given by the extension (rescaling) of the time axis
from one year (1990) to 10 years (1990-2000).
Notice the similarity between the first plot and the last one,
which is due to the invariance of seismicity under 
this transformation.
}
\label{rg_transf} 
\end{figure}

Figure \ref{rg_transf} displays the magnitude $M$ versus the occurrence time
$t$ of all worldwide earthquakes with $M \ge M_c$ for different periods of
time and $M_c$ values.
The top of the figure is for earthquakes with
$M\ge 5$
for the year 1990.
If we rise the threshold up to $M_c=6$ we get the results shown in Fig.~ 
\ref{rg_transf}(medium). 
Obviously, as there are less earthquakes in this case, 
the distribution of recurrence times 
(time interval between consecutive ``spikes'' in the plot)
becomes broader with respect to the previous case, as we know. 
The rising of the threshold can be viewed
as a mathematical transformation of the seismicity point process,
which is referred to as {\it thinning}                                    \index{thinning}
in the context of stochastic processes
\cite{Daley_Vere_Jones} 
and is also equivalent to the common {\it decimation }                    \index{decimation}
performed for
spin systems in renormalization-group transformations                     \index{renormalization-group transformation}
\cite{Kadanoff,McComb,Christensen_Moloney};
the term decimation is indeed appropriate 
as only one tenth of the events survive 
this transformation,
due to the fulfillment of the GR law with $b=1$.
Figure \ref{rg_transf}(bottom) shows the same as Fig. \ref{rg_transf}(medium) but
for ten years, 1990-1999, 
and represents a scale transformation
of seismicity (also as in the renormalization group), contracting
the time axis by a factor 10 to compensate for the previous decimation.
The similarity between Fig. \ref{rg_transf}(top) and \ref{rg_transf}(bottom)
is apparent, and is confirmed when the probability densities of
the corresponding recurrence times are calculated and rescaled following
Eq. (\ref{scaling}), see again Fig. \ref{Dtsca2}(right).

\subsection{Simple Model to Renormalize}

A simple model may illustrate these ideas 
\cite{Corral_prl.2005}.
Let us assume that seismicity could be described
as a time process for which each recurrence time
$\tau_i$ (which separates event $i-1$ and $i$) 
only depends on $M_{i-1}$, the magnitude of the 
last event that has occurred before event $i$. 
Any other dependences
are ignored, and in particular the values of the magnitudes
are generated independently from the rest of the process.
It is possible to shown that for this process the recurrence-time
density for events above $M_c'$, $D(\tau ; M_c')$, 
can be related to the recurrence-time density for events above $M_c$
conditioned to $M_{pre} \ge M_c'$ or to $M_{pre} <  M_c'$,
which we denote
$D(\tau | M_{pre} \ge M_c' ; M_c )$ and $D(\tau | M_{pre} < M_c' ; M_c )$,
respectively, where $M_{pre}$ refers to the magnitude 
of the event immediately previous to the recurrence time,
and it is assumed that $M_c' > M_c$.
The relation turns out to be
\begin{equation}
\begin{array}{lll}
     D(\tau ; M_c') & = & p D(\tau | M_{pre} \ge M_c' ; M_c )
     + qp D(\tau | M_{pre} \ge M_c' ; M_c ) \\
& * & D(\tau | M_{pre} < M_c' ; M_c )
     + q^2 p D(\tau | M_{pre} \ge M_c' ; M_c )\\ 
& * & D(\tau | M_{pre} < M_c' ; M_c )
     * D(\tau | M_{pre} < M_c' ; M_c ) + \cdots\\
& = & p D(\tau| M_{pre} \ge M_c' ; M_c )  *
\sum_{k=0}^\infty
     q^k [D(\tau | M_{pre} < M_c' ; M_c )]^{*k}
\end{array}
\label{larga}
\end{equation}
where $*$ denotes the convolution product
and $p$ is the probability that an earthquake is above $M_c'$,
given that it is above $M_c$, i.e.,
\begin{equation}
p \equiv \mbox{Prob}[M\ge M_c' | M\ge M_c]=10^{-b(M_c'-M_c)},
\end{equation}
using the GR law, whereas $q \equiv \mbox{Prob}[M < M_c' | M\ge M_c]= 1-p$.
Equation (\ref{larga}) enumerates the number of ways in which 
two consecutive events for $M\ge M_c'$ 
may be separated by a recurrence time $\tau$;
these are the number of events with $M< M_c'$ in between,
each one contributing with a probability $q$, 
and then the time $\tau$ between the two events 
is in fact a $(k+1)-$th return-time for the process with $M\ge M_c$;
from here and the independence between recurrence times
the convolutions arise.

Let us translate Eq. (\ref{larga}) to Laplace space, by using
$F(s)\equiv \int_0^\infty e^{-s\tau}F(\tau) d\tau$;
then, the convolutions turn out to be simple products, 
i.e.,
\begin{equation}
     D(s; M_c') = p D(s| M_{pre} \ge M_c' ; M_c ) 
\sum_{k=0}^\infty
     q^k [D(s | M_{pre} < M_c' ; M_c )]^k.
\end{equation}
As $q D(s | M_{pre} < M_c' ; M_c ) < 1$
the series can be summed, yielding
\begin{equation}
     D(s; M_c') 
=\frac{p D(s| M_{pre} \ge M_c' ; M_c ) }
     {1-D(s ; M_c )+
pD(s | M_{pre} \ge M_c' ; M_c )},
\end{equation}
using that $D(s ; M_c )=
pD(s | M_{pre} \ge M_c' ; M_c )+
qD(s | M_{pre} < M_c' ; M_c )$.
We have obtained an equation for the transformation 
of the recurrence-time probability density under the
thinning or decimation caused by the raising of the
magnitude threshold from $M_c$ to $M_c'$.
The second part in the process is the simple
rescaling of the distributions, to make them have
the same mean and comparable with each other;
we obtain this by removing the effect of
the decreasing of the rate, which, due to thinning,
is proportional to $p$, so,
\begin{equation}
D(\tau ; M_c') \rightarrow p^{-1} D(p^{-1}\tau ; M_c'),
\end{equation}
and in Laplace space,
\begin{equation}
D(s ; M_c') \rightarrow  D(p s ; M_c').
\end{equation}
Finally, the renormalization-group transformation       \index{renormalization-group transformation}
$\top$ is obtained by combining the decimation 
with the scale transformation, 
\begin{equation}
     \top [D(s; M_c)] =
\frac{p D(ps| M_{pre} \ge M_c' ; M_c ) }
     {1-D(ps ; M_c )+
pD(ps | M_{pre} \ge M_c' ; M_c )}.
\label{transform}
\end{equation}
A third step which is usual in renormalization-group
transformations is the renormalization of the field, 
$M$ in this case, but as we are only interested in recurrence times
it will not be necessary here.
The fixed points of the renormalization-group transformation are obtained by
the solutions of the fixed-point equation                    \index{fixed-point equation}
\begin{equation}
\top [D(s;M_c)] = D(s;M_c).
\label{fixed_point}
\end{equation}
This equation is equivalent to the scaling law for the recurrence-time densities,
Eq. (\ref{scaling}),
the only difference is that now it is expressed in Laplace space, 
as we are not able to provide the form of the operator $\top$ in real space.

\subsection{Renormalization-Group Invariance of the Poisson Process}

We can get some understanding of the transformation $\top$
by considering first the simplest possible case,
that in which there are no 
correlations in the process; so we have to 
break the statistical dependence 
between the magnitude and the subsequent
recurrence time.
This means that 
\begin{equation}
D(\tau | M_{pre} \ge M_c' ; M_c )=
D(\tau | M_{pre} <  M_c' ; M_c )=
D(\tau ; M_c ) \equiv D_0(\tau ; M_c ) 
\end{equation}
and then the renormalization transformation
turns out to be
\begin{equation}
   \top [D_0(s; M_c )] =
   \frac{p D_0 (ps; M_c )}{1- q D_0 (ps; M_c )}.
\end{equation}
if we introduce $\omega \equiv p s$ and substitute $p=\omega/s$
and $q=1-\omega/s$ in the fixed-point equation
$\top D_0(s; M_c ) = D_0(s; M_c )$,
we get, separating variables and equaling to an arbitrary constant 
$k$
\begin{equation}
\frac 1 {sD_0(s; M_c )} - \frac 1 {s} =
\frac 1 {\omega D_0(\omega; M_c )} - \frac 1 {\omega} \equiv  k;
\end{equation}
due to the fact that $p$ and $s$ are independent variables 
and so are $s$ and $\omega$.
The solution is then
\begin{equation}
D_0(s; M_c )={(1+ ks)^{-1}},
\label{PoissonLaplace}
\end{equation}
which is the Laplace transform of an exponential
distribution,
\begin{equation}
 D_0(\tau; M_c )=k^{-1} e^{-\tau /k}.
\end{equation}
The dependence on $M_c$ enters by means of $k$,
as $k=\langle \tau (M_c)\rangle$;
in the case of seismicity the GR law holds and 
$k= R^{-1}(M_c)=R_0^{-1} 10^{b M_c}$.

Summarizing, we have shown that the only process 
without correlations 
which is invariant under a renormalization-group 
transformation of the kind we are dealing with
is the Poisson process. 
This means that if one considers as a model of seismicity
a renewal process                                          \index{renewal process}
(i.e., independent identically distributed return times)
with uncorrelated magnitudes, then
the recurrence-time distributions 
will not verify a scaling law 
when the threshold $M_c$ is raised,
except if $D(\tau;M_c)$ is an exponential
(which constitutes the trivial case of a Poisson process).

Even further, the Poisson process is not only a fixed point
of the transformation, but a stable one (or attractor)
for a thinning transformation in which 
events are randomly removed from the process (random thinning).
If magnitudes are assigned to any event independently
of any other variable (other magnitudes or recurrence times)
the decimation of events after the risen of the threshold
$M_c$ is equivalent to a random thinning,
and therefore 
the resulting process must converge to a Poisson process, 
under certain conditions \cite{Daley_Vere_Jones}.

The fact that for real seismicity the scaling function $f$ is not
an exponential tells us that our renormalization-group transformation
is not performing a random thinning; this means that the magnitudes 
are not assigned independently on the rest of the process
and therefore there exists correlations in seismicity.
This of course is not new, but let us stress that
correlations are fundamental for the existence of the scaling 
law (\ref{scaling}):
the only way to depart from the trivial Poisson process is to consider 
correlations between recurrence times and magnitudes in the process.
This is the motivation for the model explained in this section,
for which we have chosen the simplest form of correlations
between magnitudes and subsequent recurrence times.
In fact, Molchan has shown that even for this correlated model 
the Poisson process is the only possible fixed point,
implying that this type of correlations are too weak
and one needs a stronger dependence of the recurrence times
on history to depart from the Poisson case.
After all, this is not surprising, as we know 
from the study of equilibrium critical phenomena 
that in order to flow away
from trivial fixed points, long-range correlations are
necessary. 
Therefore, the problem of finding a model of correlations
in seismicity yielding a nontrivial recurrence-time scaling
law is open.


\section{Correlations in Seismicity}

In the preceding section we have argued that 
the existence of a scaling law for recurrence time distributions
is inextricably linked with the existence of correlations in the process,
in such a way that correlations determine the form 
of the recurrence-time distribution.
In consequence, an in-depth investigation of correlations
in seismicity is necessary.

Our analysis will be based in the conditional probability density;
for instance, for the recurrence time we have,
$$
     D(\tau | X) \equiv  \frac{\mbox{Prob}[\tau < \mbox{ recurrence time } 
     \le \tau + d\tau \,  |\, X]}{d\tau},
$$
where $|X$ means that the probability is only computed for the cases
in which the condition $X$ is fulfilled. If it turns out to be that
$D(\tau | X) $ is undistinguishable from the unconditional density,
$D(\tau) $, then, the recurrence time is independent on the condition $X$;
on the contrary, if both distributions turn out to be significantly
different, this means that the recurrence time depends on the condition $X$
and we could define a correlation coefficient to account for this
dependence, although in general we might be dealing with a nonlinear correlation.
\index{correlation}

Moreover, as we will compare values of the variables in different times
(for example, the dependence of the recurrence time $\tau_i$ on the value of the 
preceding recurrence time, $\tau_{i-1}$, for all $i$),
we introduce a slight modification in the notation, particularly
with respect the previous section, including the subindices
denoting the ordering of the events
in the probability distributions.
Further, in order to avoid complications 
in the notation, we will drop the dependence
of the conditional density on $M_c$ when
unnecessary.

\subsection{Correlations between recurrence times}

Let us start 
with the temporal sequence of occurrences,
for which we obtain the conditional 
distributions $D(\tau_i | \tau_a \le \tau_{i-1} < \tau_b )$;
in particular we distinguish two cases; short preceding recurrence times,
$D(\tau_i | \tau_{i-1} < \tau_b )$, where $\tau_b$ is small, and
long preceding recurrences, $D(\tau_i | \tau_{i-1} \ge \tau_a  )$,
with $\tau_a$ large.
The results, both for worldwide seismicity and for Southern-California
stationary seismicity, turn out to be practically the same, 
see Fig. 6 of Ref. \cite{Corral_tectono}.
For short $\tau_{i-1}$,
a relative increase in the number of short $\tau_i $ 
and a decrease of long $\tau_i $ is obtained, in comparison
with the unconditional distribution, which leads to a steeper 
power-law decay of the conditional density for short and
intermediate times. 
In the opposite case, long $\tau_{i-1}$'s imply a decrease in the number
of short $\tau_i $ and an increase in the longer ones, in such a way that
a flatter power-law exists here.
In any case, the behavior for long $\tau_i$ is exponential.
So, short $\tau_{i-1}$'s imply an average reduction
of $\tau_i $ and the opposite for long $\tau_{i-1}$'s,
and then both variables are positively correlated.


This behavior corresponds to a clustering of events,
in which short recurrence times tend to be close to 
each other, forming clusters of events, while longer times 
tend also to be next each other. This clustering effect is 
different from the clustering reported in previous sections, associated
to the non-exponential nature of the recurrence-time distribution,
but is similar, in some sense, to the usual clustering of aftershock sequences,
as these sequences also show this kind of correlations,
although mainly due to the time-variable rate.

In fact, the case of nonstationary seismicity was
studied by Livina et al. for Southern California \cite{Livina,Livina2},
with the same qualitative behavior.
These authors explain their results in terms of the persistence
of the recurrence time, which is a concept equivalent to 
the kind of clustering we have described. 
The results could be also similar
to the long-term persistence observed in climate records.

The effect of correlations can be described in terms of a scaling
law, which constitutes a generalization of the scaling law
for (unconditioned) recurrence time distributions, Eq. (\ref{scaling}).
In this way, we can write the conditional recurrence-time distribution
in terms of a scaling function which depends on two variables, 
$R\tau_i$ and $R\tau_{a}$, or $R\tau_i$ and $R\tau_{b}$,
see Ref.  \cite{Livina2}.  
Further, the study of correlations between recurrence times can
be studied beyond consecutive events, i.e., we can measure the
distribution of $\tau_i$ conditioned to $\tau_{i-2}$, 
or $\tau_{i-3}$, etc. The results for these distributions
show no qualitative difference, at least up to $i-10$, 
in comparison with what we have explained for $i-1$.

The main results of this subsection can be summarized
in one single sentence, reflecting the positive correlation
between recurrence times: {\it the shortest the time 
between the two last earthquakes, the shortest the 
recurrence of the next one,} on average.

\subsection{Correlations between recurrence time and magnitude}

If magnitudes
are taken into account, there are two main types of correlations
with the recurrence times. First, we consider how the magnitude
of one event influences the recurrence time of a future event,
in particular the next one,
measuring $D(\tau_i | M_{i-1} \ge M_c')$.
The results for the case of worldwide seismicity and for
Southern California (in a stationary case) are again 
similar and show a clear (negative) correlation between $M_{i-1}$
and $\tau_i $ \cite{Corral_comment}.

\begin{figure}
\centering
\includegraphics[height=4.1cm]{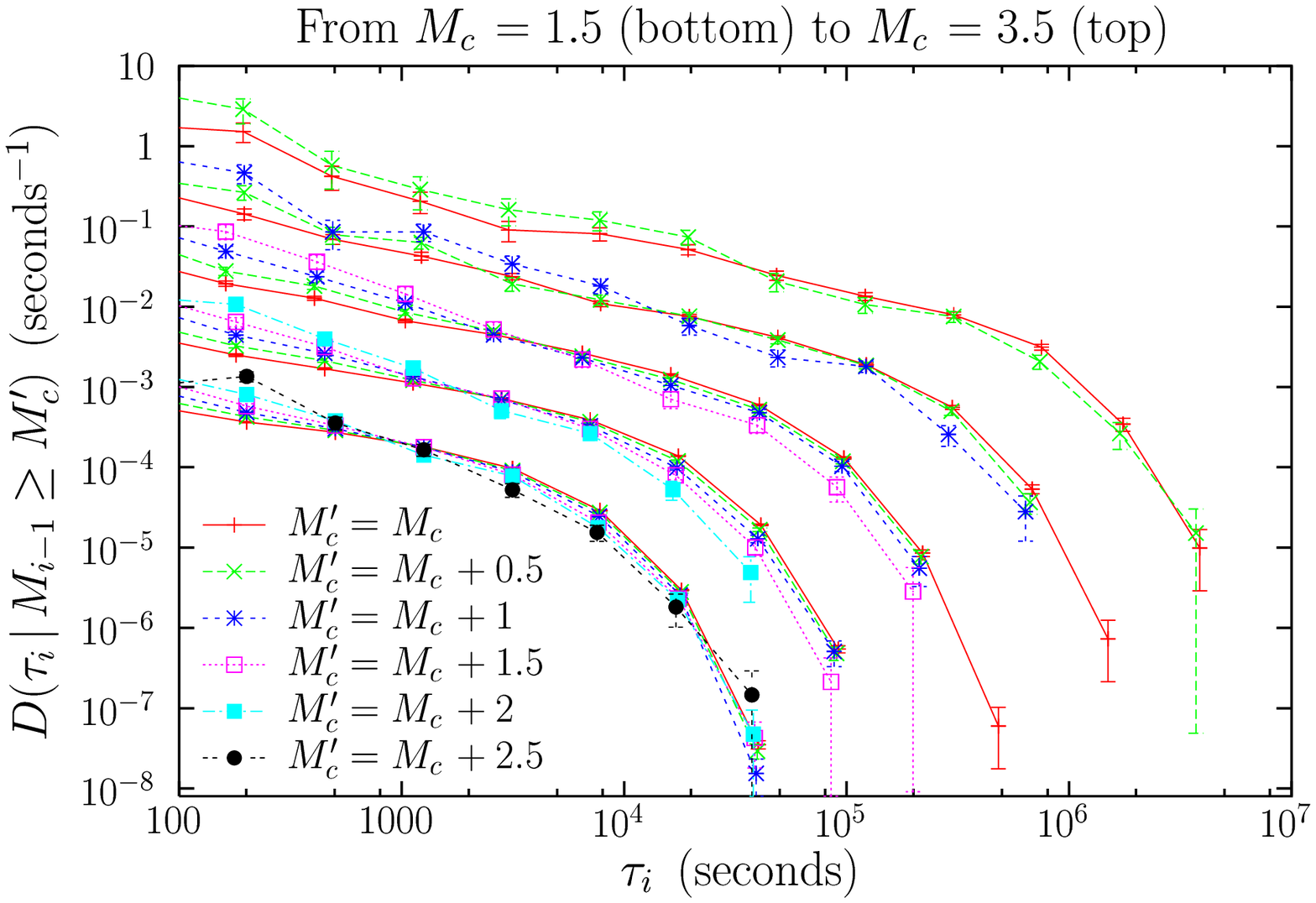}
\includegraphics[height=4.1cm]{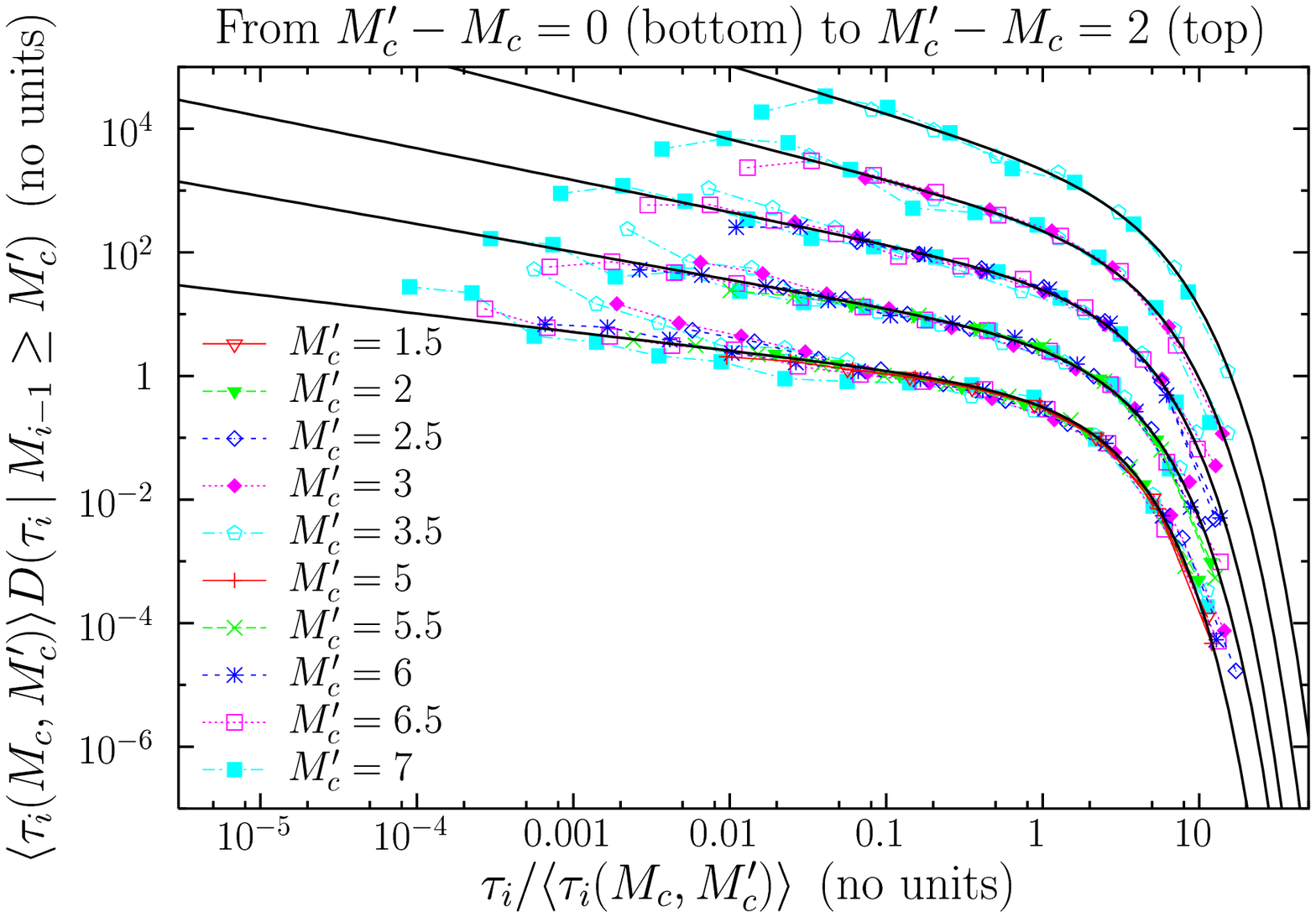}
\caption{
(left)
Recurrence-time distributions for Southern-California 
conditioned to the value of the preceding magnitude,
for the stationary period comprised between 1994-1999.
Each set of curves, which have been shifted up for clarity sake,
corresponds to a value of $M_c$, which is, 
from bottom to top: 1.5, 2, 2.5, 3, 3.5.
(right)
Rescaled recurrence-time distributions conditioned to the value
of the preceding magnitude, both for worldwide and for Southern-California
stationary seismicity. Each set of data, shifted up again for clarity sake,
corresponds this time to a different value of $M_c'-M_c$, 
these being 0, 0.5, 1, 1.5, and 2, from bottom to top,
and is fit by a gamma distribution, with (decreasing) power-law
exponents 0.30, 0.45, 0.52, 0.65, and 0.77, respectively.
}
\label{Dtaumpre} 
\end{figure}

Figure \ref{Dtaumpre}(left) shows, for Southern-California
in the stationary case, how larger values of the 
preceding magnitudes, given by $M_{i-1} \ge M_c'$,
lead to a relative increase in the number of short $\tau_i $ 
and a decrease in long $\tau_i $, implying that 
$M_{i-1}$ and $\tau_i $ are anticorrelated.
For the cases for which the
statistics is better, the densities show the behavior
typical of the gamma distribution, with a power law 
that becomes steeper for larger $M_c'$;
the different values of the power-law exponent are given 
at the figure caption.

Remarkably, this behavior can be described by a scaling law for which
the scaling function depends now on the difference between
the threshold magnitude for the $i-1$ event, $M_c'$, and the threshold
for $i$, $M_c$, i.e.,
$$
D(\tau_i | M_{i-1} \ge M_c'; M_c) = R(M_c,M_c') f(R(M_c,M_c')\tau_i , M_c'-M_c)
$$
with $R(M_c,M_c')\equiv 1/\langle\tau(M_c,M_c')\rangle$
and 
$\langle\tau_i(M_c,M_c')\rangle$ the mean of the distribution
$D(\tau_i | M_{i-1} \ge M_c';M_c)$. 
The original scaling law for unconditioned distributions,
Eq. (\ref{scaling}), is recovered taking the case $M_c'=M_c$.
Figure \ref{Dtaumpre}(right) illustrates this scaling law
both for worldwide and for Southern-California stationary seismicity.

\begin{figure}
\centering
\includegraphics[height=4.1cm]{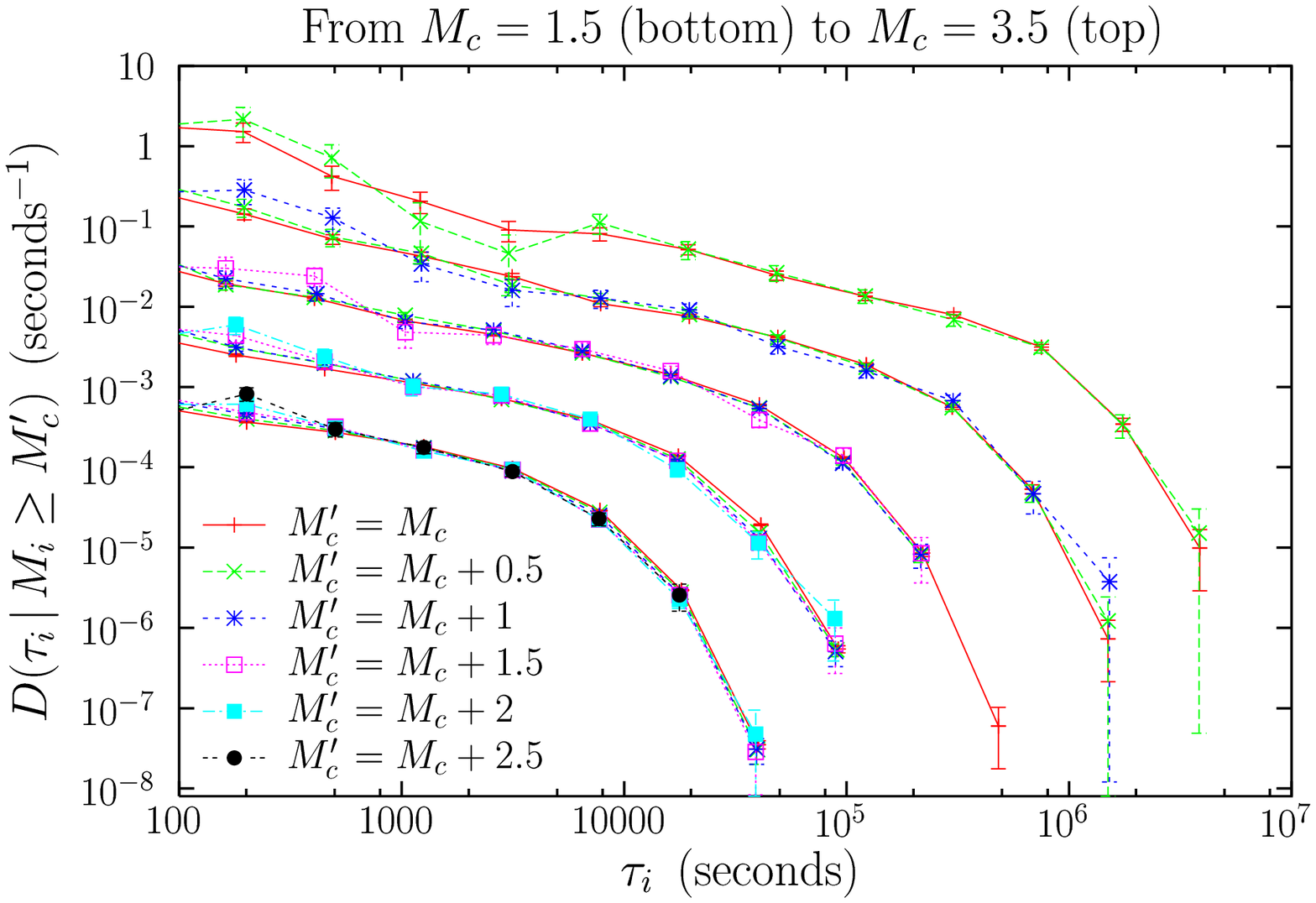}
\includegraphics[height=4.1cm]{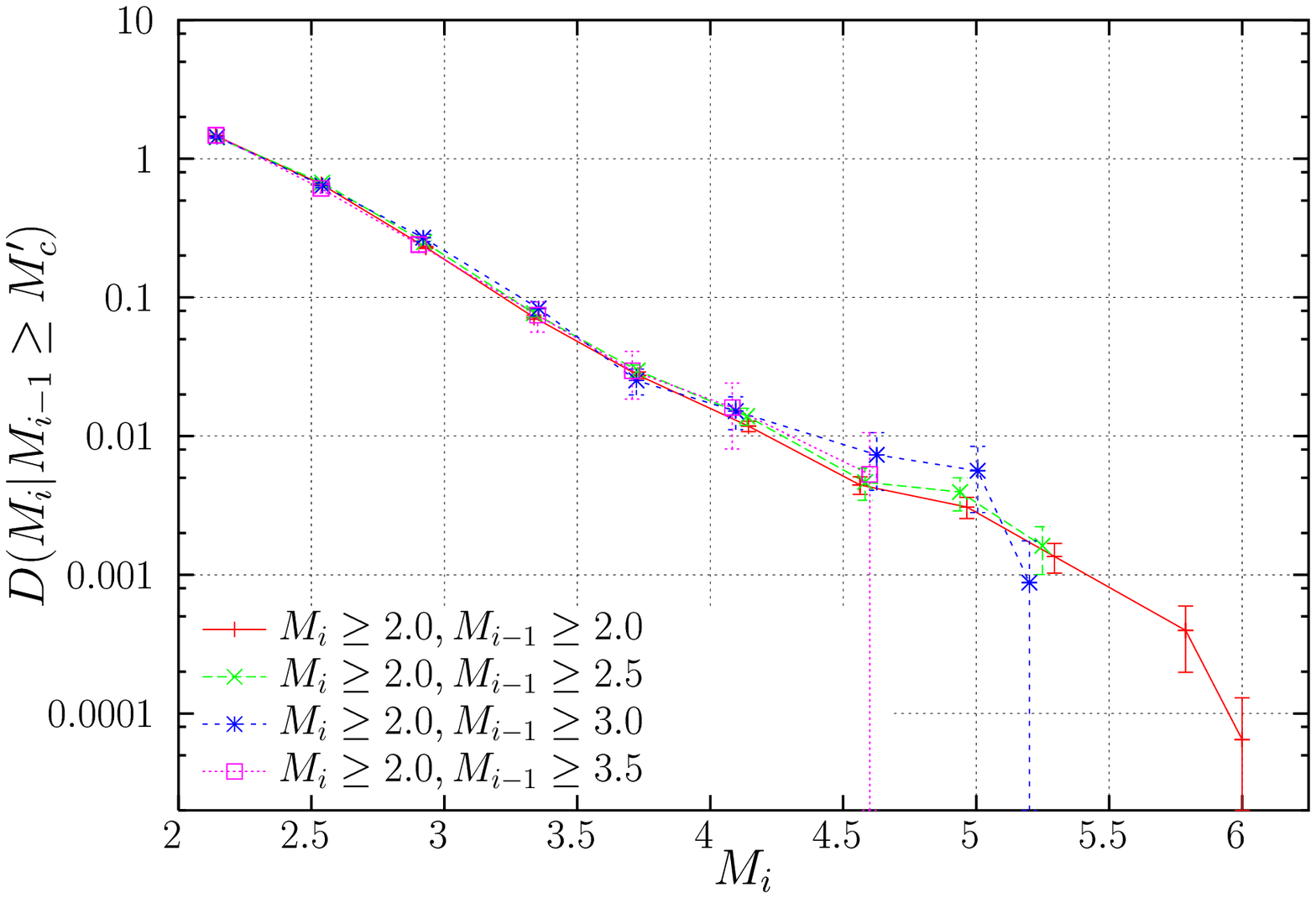}
\caption{
(left) Recurrence-time distributions for Southern-California 
conditioned to the value of the magnitude of the incoming event,
for the stationary period comprised between 1994-1999.
The curves have been shifted up and each set 
corresponds to a value of $M_c$, which ranges
from 1.5 to 3.5 in steps of 0.5, from bottom to top.
(right)
Distribution of magnitudes conditioned to the 
value of the preceding magnitude for Southern California
for the set of stationary periods indicated at the text.
Only events separated by $\tau \ge 30$ min have been 
considered.
\label{Dtaum}
}
\end{figure}

The second kind of correlations deals with
how the recurrence time to one event $\tau_i$
influences its magnitude $M_i$, or, equivalently, how
the recurrence time to one event depends on the magnitude
of this event. 
For this purpose, we measure $D(\tau_i | M_i \ge M_c')$;
the results for Southern-California stationary seismicity 
are shown in Fig. \ref{Dtaum}(left).
It is clear that in most of their range the distributions
are nearly identical, and when some difference is present
this is inside the uncertainty given by the error bars.
However, there is one exception: 
very short times, $\tau_i \simeq 3$ min, 
seem to be favored by larger magnitudes, or, in other words,
short times lead to larger events;
nevertheless, due to the short value of the time involve,
we can ignore this effect. Then,
the recurrence time and the magnitude after it
can be considered
as independent from a statistical point of view,
with our present resolution 
(it might be that a very weak dependence is hidden
in the error bars of the distributions).

Further, as in the previous subsection, we have gone several more steps
backwards in time, measuring conditional distributions 
up to $D(\tau_i | M_{i-10} \ge M_c')$, and also we have 
extended the conditional distributions to the future, 
measuring $D(\tau_i | M_{j} \ge M_c')$ with $j>i$.
The results are not qualitatively different than
what is described previously,
with the first type of distributions dependent on $M_c'$
and the second type independent.
From here we can conclude the dependence of the recurrence time   \index{dependence of the recurrence time with history}
and the independence of the magnitude with the 
sequence of previous recurrence times, at least with our present
statistics and resolution.

Two sentences may serve to summarize the behavior of seismicity
described here. First,
{\it the bigger the size of an earthquake, 
the shortest the time till next}, 
due to the anticorrelation between magnitudes and forward recurrence
times. Note that in the case of stationary seismicity
this result is not trivially derived from the law
of aftershock productivity.
Second, the belief that {\it the longer the recurrence time for an earthquake,
the bigger its size}, is false, as the magnitude is uncorrelated
with the previous recurrence times.
This shows clearly the time irreversibility of seismicity.     \index{time irreversibility of seismicity}

\subsection{Correlations between magnitudes \label{mm}}

Although not directly related with the temporal properties,
we study the correlations between consecutive magnitudes,
$M_{i-1}$ and $M_i$, by means of the distribution $D(M_i | M_{i-1} \ge M_c')$.
As the analysis of the 1994-1999 period for Southern California
did not provide enough statistics for the largest events,
we considered a set of stationary periods, these being:
Jan 1, 1984 - Oct 15, 1984; 
Oct 15, 1986 - Oct 15, 1987;
Jan 1, 1988 - Mar 15, 1992; 
Mar 15, 1994 - Sep 15, 1999; and
Jul 1, 2000 - Jul 1, 2001;
all of them visible in Fig. \ref{Nt}.

Again we find that both worldwide seismicity and stationary 
Southern-California seismicity share the same properties,
but with a divergence for short times.
Figure 10 of Ref. \cite{Corral_tectono} shows the distributions corresponding
to the two regions, and whereas for the worldwide case
the differences in the distributions for different $M_c'$
are compatible with their error bars,
for the California case there is a systematic deviation, implying
a possible correlation.

In order to find the origin of this discrepancy we include an extra condition,
which is to restrict the events to the case of large enough recurrence times,
so we impose $\tau_i \ge 30$ min. 
In this case, the differences in Californian
distributions 
become no significant, see Fig. \ref{Dtaum}(right), 
which means that the significant correlations between consecutive magnitudes
are restricted to short recurrence times \cite{Corral_comment}. 
Therefore, we conclude that the Gutenberg-Richter law is valid independently
of the value of the preceding magnitude, provided that short times are not
considered.
This is in agreement with the usual assumption in the ETAS model,
in which magnitudes are generated from the Gutenberg-Richter distribution
with total independence of the rest of the process
\cite{Ogata99,Helmstetter_pre02}.
Of course, this independence is established within the errors
associated to our finite sample. It could be that the dependence
between the magnitudes is weak enough for that 
the changes in distribution are not larger than the
uncertainty. With our analysis, only a much larger data set 
could unmask this hypothetical dependence.

The deviations for short times may be an artifact 
due to the incompleteness of earthquake catalogs at
short time scales, for which small events are not recorded.
Helmstetter et al. \cite{Helmstetter_Kagan_forecast} propose a formula for 
the magnitude of completeness in Southern California
as a function of the elapsed time since
a mainshock and its magnitude; applying it to our stationary periods,
for which the larger earthquakes have magnitudes ranging from 5 to 6,
we obtain that a time of about 5 hours is necessary in order that the magnitude of 
completeness reaches a value below 2 after a mainshock of magnitude 6.
After mainshocks of magnitude 5.5 and 5 this time reduces to about
1 hour and 15 min, extrapolating Helmstetter et al.'s
results. 
In any case, it is perfectly possible that large mainshocks 
(not necessarily the preceding event) 
induce the loss of small events in the record and
are the responsible of the deviations
from the Gutenberg-Richter law at small magnitudes
for short times.
If an additional physical mechanism is behind this behavior,
this is a question that cannot be answered  with this kind of
analysis.

As in the previous subsection, we have performed measurements 
of the conditional distributions for worldwide seismicity
involving different $M_i$ and
$M_j$, separated up to 10 events, with no significant variations
in the distributions, as expected. 
This confirms the independence
of the magnitude $M_i$ with its own history.                     
\index{independence of magnitude with history}

These results, together with those of the previous
subsection allow to state that 
{\it an earthquake does not ``know'' how big is going to be}
(at least from the information recorded at the catalogs, 
disregarding spatial structure, and with our present resolution)
\cite{Corral_comment,Corral_tectono}.

\subsection{Correlations between recurrence times and distances}

Up to now we have considered seismicity as a point process in time,
marked by the magnitude. 
If, in addition to this, we take into account the spatial degrees of freedom,
these new variables allow to study other types of correlations.
Of outstanding importance will be the distances between earthquakes,
and specially the distances between consecutive earthquakes, 
which we may call jumps.

The correlations between jumps and recurrence times
have been investigated in Ref. \cite{Corral_distances},
and they show a curious behavior. There are two kinds
of recurrence time distributions conditioned to the distance, 
one for short distances, which can be represented by a gamma
distribution with a decaying power law of exponent around 0.8,
and the distribution for long distances, which is an exponential.
It is clear then that in one case we are dealing with aftershocks
and in the other with Poissonian events.
The particularity of these distributions is that they are independent
on the distances, provided that the set of values of the distances
are short, or long. For worldwide earthquakes, the difference between 
short and long distances is around $2^\circ$ (200 km),
whereas for Southern California this value is $0.1^\circ$,
approximately.

We may note that
the (unconditional) distribution of recurrence times is then a mixture
of these two kind of conditional distributions, and therefore, 
the existence of a universal recurrence-time distribution is a
consequence of a constant proportion of short and long distances
in seismicity, or of aftershocks and uncorrelated events.

\section{Bak et al.'s Unified Scaling Law}

Bak, Christensen, Danon, and Scanlon introduced 
a different way to study recurrence times in earthquakes
\cite{Bak.2002}.
They divided the region of Southern California into approximately 
equally-sized squared subregions (when longitude and latitude 
are taken as rectangular coordinates), and computed
the series of recurrence times for each subregion.
The main difference with the procedure explained in
the previous sections is that
Bak et al. included all the series
of recurrence times into a unique recurrence-time
distribution, performing therefore a mixing
of the distributions for all subregions.
As seismic rate displays large variations in space
(compare the rates of occurrence in Tokyo and in Moscow,
and the same happens at smaller scales)
Bak et al.'s procedure leads to a very broad distribution 
of recurrence times.

\begin{figure}
\centering
\includegraphics[height=4.1cm]{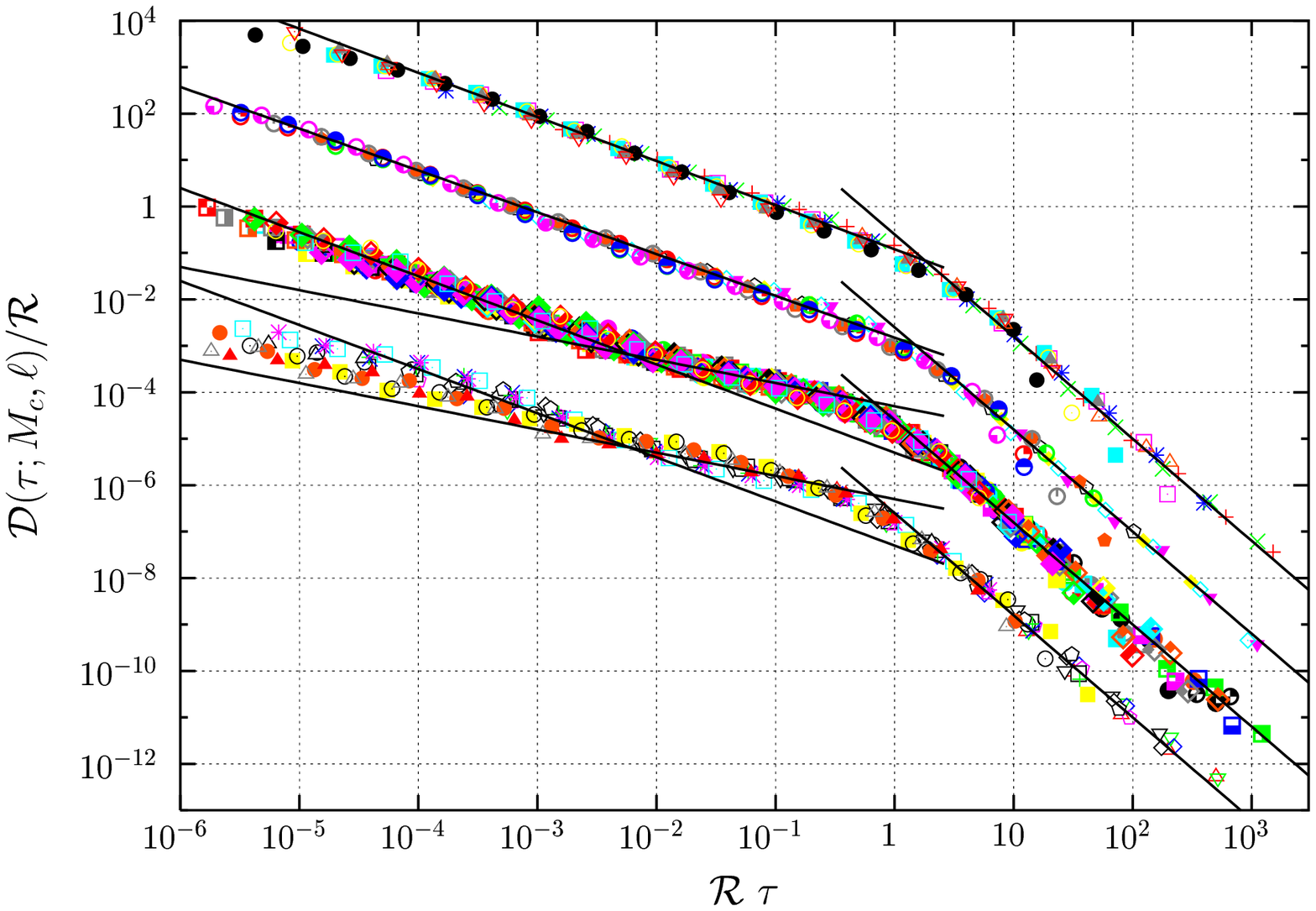}
\includegraphics[height=4.1cm]{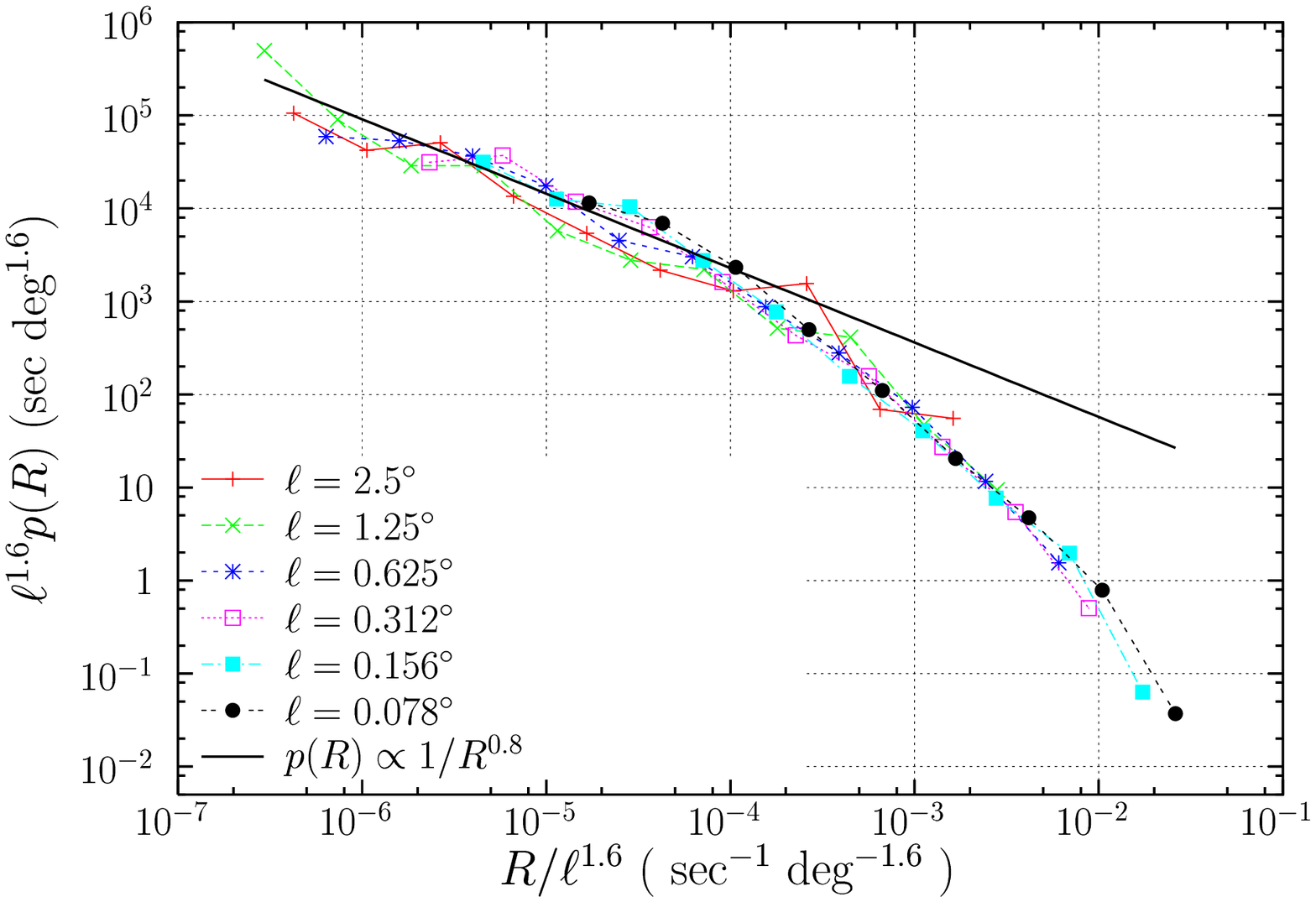}
\caption{
(left) Mixed recurrence-time densities (defined in the way of Bak et al.)
and rescaled by $\mathcal{R}$, for several values of $\ell$ and $M_c$.
The different sets of curves correspond, from top to bottom, to 
(i) Southern California, 1984--2001;
(ii) Northern California, 1985--2003;
(iii) Stationary seismicity: Southern California, 1988--1991; worldwide, 1973--2002;
Japan, 1995--1998; and Spain, 1993--1997; 
(iv) Stationary seismicity: New Zealand, 1996--2001, and New Madrid, 1975--2002;
A total of 84 distributions are shown, $\ell$ ranging from $0.039^\circ$  to $45^\circ$, 
and $1.5 \le M_c  \le 6$. 
The distributions are shifted to the bottom for clarity sake.
All the left tails are fit by a (decreasing) power-law with exponent 2.2,  
the right part of the distributions
are fit by a power law with exponent 0.95 or 0.9,
see Ref. \cite{Corral_physA.2004}.
(right)
Distribution of mean rates of occurrence for events with $M\ge 2$ in Southern-California, 
1984--2001 (averaging 1984--1992 and 1993-2001), using diverse values of $\ell$.
The distributions are rescaled by $L^{1.6}$ and the straight line is a power law 
fit which turns out to be $\propto 1/R^{0.8}$.
}
\label{hetero}       
\end{figure}

It was found that the recurrence-time densities
defined in this way,
$\mathcal{D}(\tau;M_c,\ell)$,
for different magnitude thresholds $M_c$ and different
linear size $\ell$ of the subregions, verify the following
scaling law (see Fig. \ref{hetero}(left)),
$$
\mathcal{D}(\tau;M_c,\ell) = \mathcal{R} F (\mathcal{R} \tau),
$$
which was named {\it unified scaling law},                   \index{unified scaling law}
where 
$F$ is the scaling function, showing
a power-law decay with exponent close to 1 
for small recurrence times
and a different
power-law decay for long times, with exponent around 2.2
\cite{Corral_pre.2003},
whereas
$\mathcal{R}(M_c,\ell)$ is 
the spatial average of the mean seismic rate, i.e., 
the average of $R_{xy}(M_c,\ell)$
for all the regions with seismic activity (labeled by $xy$),
so,
%
${\mathcal{R}}={\sum_{xy} R_{xy}}/{n}$,
%
where ${n}$ is the number of such regions.
Note that $\mathcal{R}$ is the inverse of the mean 
of $\mathcal{D}$.
From the GR law for each region, $R_{xy} = R_{xy0} 10^{-b M_c}$ 
and 
from the fractal scaling of ${n}$ with $\ell$,
${n}=({\mathcal{L}}/\ell)^{d_f}$, we get,
${\mathcal{R}}=R_0 (\ell/{\mathcal{L}})^{d_f} 10^{-b M_c}$,
with $R_0(\mathcal{L})=\sum_{xy} R_{xy0}(\ell)$ 
and ${\mathcal{L}}$ a rough measure
of the linear size of the total area under study.
Therefore we can write the scaling law as
$$
\mathcal{D}(\tau;M_c,\ell) = \ell^{d_f} 10^{-b M_c}\tilde 
F (\ell^{d_f} 10^{-b M_c} \tau),
$$
which relates the recurrence-time density, defined
in the Bak et al.'s way, with the GR law and with 
the fractal distribution of epicenters,
and from here the name of unified scaling law.
Molchan and Kronrod have studied this law
in the framework of multifractals \cite{Molchan05}.

Later it was found that the unified scaling law holds beyond 
the case of Southern California, for instance for 
Japan, Spain, New Zealand, New Madrid (USA), or Iceland,
as well as worldwide
\cite{Corral_physA.2004,Davidsen_Goltz}.
However,
it turned out that the scaling function is not universal,
as there are differences for different regions, 
mainly in the crossover between short and long times, although the 
value of the long-time power-law exponent seems
to be in all cases 2.2, and therefore universal,
see Fig. \ref{hetero}(left).
The deviations from the hyperbolic-like behavior
(exponent close to one) 
for very short times have also been studied 
\cite{Davidsen_Goltz}.

It is clear that the short-time exponent must be related (but not identical!)
to the Omori $p-$value;
on the other hand, the long-time exponent is a consequence of
a power law distribution of seismic rates in space, as we now show
\cite{Corral_Christensen}.
Therefore, 
with the purpose of understanding the relation of Bak et al.'s
results with the rest of this work, let us generalize
the nonhomogeneous Poisson-Omori process previously introduced,
in order to include the universal scaling law for Omori sequences.
We have explained that 
these sequences can be characterized by
an $r-$dependent recurrence-time probability density of the form
$D(\tau |r) \propto r^\gamma \tau^{\gamma-1}e^{-r\tau /a}$
(note that this includes the nonhomogeneous Poisson process,       \index{nonhomogeneous Poisson process}
given by $\gamma=a=1$, but for real field data $\gamma\simeq 0.7$).
We expect that, for a given spatial area, 
this is valid not only for Omori sequences
but also for a general time-varying rate; then
the overall probability density of the recurrence times,
independently of $r$,
is given by the mixing of all $D(\tau|r)$ \cite{Corral_pre.2003},
\begin{equation}
D(\tau | r_m) = \frac{1}{\mu} \int_{r_m}^{r_0} r D(\tau | r) \rho(r) dr, 
\end{equation}
in fact, this is just Eq. (\ref{mixing}); 
we recall that $\rho(r)$ is the density of rates, 
$\mu$ is the mean rate,
$r_0$ is the maximum rate,
and
$r_m$ is the minimum rate,
related to the background seismicity level.
Note that we have emphasized the dependence of
the resulting distribution on $r_m$.

Let us consider that the distribution of rates 
comes essentially from Omori sequences, 
then, as we already know, 
$\rho(r)= C/r^{1+1/p}$.
The analysis is simplified for 
$\gamma=1/p$, although the
conclusions will be of general validity,
so, in this case, 
\begin{equation}
D(\tau|r_m) \propto \frac {C}{ \mu} \, 
\frac{(e^{-r_m \tau /a}-e^{-r_0 \tau/a})}{ \tau^{2-1/p}},
\end{equation}
where, in the same way as for a nonhomegeneous Poisson process,  
the minimum rate $r_m$ determines the exponential tail
of $D(\tau|r_m)$ for large $\tau$, which is
preceded by a decreasing power law with exponent $2-1/p$ if $r_0 \gg r_m$.

Up to now we have arrived to a slightly different, more convenient
variation of the distribution corresponding to a nonhomogeneous
Poisson process.
Next step is to take into account the spatial 
degrees of freedom, fundamental in Bak {et al.}'s approach.
In fact, as we have explained, their approach performs a mixing 
of recurrence times coming from different
spatial areas (or subregions), which are characterized by 
disparate seismic rates.
In particular, each area will have a different
$r_m$, depending on its background seismicity level.
As the minimum rate is difficult to measure
(it depends on the size of the time intervals selected),
we assume that the minimum rate $r_m$ 
is somehow proportional to the mean rate of the sequence $\mu$,
which in turn is in correspondence with the mean rate
in the area, $R$.
This spatial heterogeneity of seismicity                 \index{seismicity spatial heterogeneity}
can be well described by a 
power-law probability density of mean rates $R$,         \index{distribution of mean seismic rates}
$p(R)\propto 1/R^{1-\alpha}$,
with $\alpha\simeq 0.2$,
see Fig. \ref{hetero}(right) and Ref. \cite{Corral_pre.2003};
then,
$$
p(r_m)\propto 1/r_m^{1-\alpha}
$$
                                                         \index{distribution of minimum seismic rates}
and therefore the recurrence-time probability density comes from
the mixing,
\begin{equation}
\mathcal{D}(\tau) \propto \int_{r_{mm}}^{r_{mM}} r_m D(\tau|r_m) p(r_m)dr_m
\end{equation}
where $r_m$ varies between $r_{mm}$ and $r_{mM}$.
Integration, taking into account that $C/\mu$ depends on $r_m$,
leads, for $ r_{mm}\tau \ll 1 \ll r_{mM}\tau$,
to
\begin{equation}
\mathcal{D}(\tau) \propto
1 / {\tau^{2+\alpha}} 
\label{2ndpower}
\end{equation}
In this way the power law for long times, 
reflects the spatial distribution of rates.
The universal value of the exponent $2+\alpha$
\cite{Corral_physA.2004},
would imply the universality of seismicity spatial heterogeneities. \index{seismicity spatial heterogeneity}
In consequence, Bak {et al.}'s unified scaling law provides a
way to measure these properties.
Further, Eq. (\ref{2ndpower}) shows that the change of exponent
in $\mathcal{D}(\tau)$ appears
for $\tau$ larger than $1/r_{mM}$, which corresponds,
for the area of highest seismicity, to the
mean of events that are in the tail of the Omori sequence,
or in background seismicity, and therefore at the
onset of correlation with the mainshock.
It is in this sense that the change of exponent separates
events with different correlation.
On the other hand, the power law for short times
is not affected by the spatial mixing and therefore
$\mathcal{D}\propto 1/\tau^{2-1/p}$.


\section{Conclusions}

We hope we have convinced the reader about the interest to
study of the temporal features of seismicity. 
The research of the author was illuminated by the ideas and
philosophy of the late Per Bak.







\printindex
\end{document}